\documentclass{aa}  

\usepackage{graphicx}
\usepackage{amsmath}
\usepackage{tikz-network}
\usepackage{supertabular,booktabs}
\usepackage{multirow}
\usepackage[normalem]{ulem}
\usepackage{hyperref}
\usepackage{amsmath}

\definecolor{darkgreen}{rgb}{0.0,0.75,0.0}

\def\mwfid{\texttt{mw-fid}}
\def\dwfid{\texttt{dwarf-fid}}
\def\flexce{\texttt{flexCE}}
\usepackage{txfonts}
\begin{document}

   \title{Disentangling chemical evolution histories with phylogenetic trees}

   \author{R. Canales\inst{1} \and P.  Jofré\inst{2} \and C. Aguilera-Gómez\inst{1} \and P. Das\inst{3} \and R. Yates\inst{4} \and T. Signor\inst{2,5} \and
   P. Tissera\inst{1} \and X. Hua\inst{6} \and R. Foley\inst{7}}

   \institute{Instituto de Astrofísica, Pontificia Universidad Católica de Chile, Av. Vicuña Mackenna 4860, 782-0436 Macul, Santiago, Chile \\ 
              \email{rebeca.canales@uc.cl}
        \and
          Instituto de Estudios Astrofísicos, Universidad Diego Portales, Av. Ejército Libertador 441, Santiago, Chile \\
              \email{paula.jofre@mail.udp.cl}
          \and 
          School of Mathematics and Physics, University of Surrey, Guildford, Surrey, GU2 7XH, UK
          \and
          Centre for Astrophysics Research, University of Hertfordshire, Hatfield, AL10 9AB, UK
          \and
          Inria Chile Research Center, Av. Apoquindo 2827, piso 12, Las Condes, Santiago, Chile
          \and
          Mathematical Sciences Institute, Australian National University, Canberra ACT 2601, Australia
          \and 
          Leverhulme Centre for Human Evolutionary Studies, Department of Archaeology, University of Cambridge, Fitzwilliam Street,
Cambridge CB2 1QH, UK
    }

   \date{}

  \abstract
   {Chemical abundances encode the fossil record of galaxy evolution in a complex and diverse way that requires innovative approaches to reconstruct galactic histories. }
   {We investigate the power of using phylogenetic methods to disentangle different evolutionary pathways in analytical chemical evolution models.}
   {We ran 1024 one-zone chemical evolution models using \texttt{flexCE}. The resulting chemical abundances were then combined with those of two fiducial models, \mwfid\ and \dwfid. These combined abundances were then used to determine which combinations produce two-branched phylogenetic trees and how purely these trees split the two input models. We used random forests and Shapley analysis to predict which model combinations return well-separated trees and explain which input parameters are most important for this. We also studied the abundance patterns as well as star formation rates, mass accumulation, and branch lengths. }
   {We find that $\eta$, the mass-loading outflow parameter in \texttt{flexCE}, had the largest impact in separating models into separate branches, due to its importance in driving the chemical enrichment rates and total abundances. Star formation rates and mass accumulation had some impact on $\eta$, but no direct relation between these quantities and the abundance planes was found. We also found that the branches in our trees connected through the most metal-rich tips, which is opposite to how phylogenetic trees connect in biological systems. This opens new avenues of phylogenetic research in different fields. }
  {Phylogenetic trees help reconstruct histories when there is information that is inherited between generations, which is the case of the chemical elements in galaxy evolution. Branch topologies can provide information about the rates of evolutionary change of the various populations, and the connection between branches also contains information about their shared history. This work brings us a step forward in understanding galaxy evolution through cross-disciplinary research. }


   \maketitle

\section{Introduction}

The cornerstone of Darwin’s theory of evolution was that of ``descent with modification,’’ wherein the traits of organisms are passed down from ancestor to descendant, from parent to offspring, with alteration between generations. This concept produces the ``tree of life,’’ through which all organisms can be linked by descent back to a series of common ancestors \citep{darwin1859}. Since being introduced, phylogenetic techniques have been developed to reconstruct such trees with new data 
\citep{bromham2016introduction, felsenstein1985phylogenies, yang2014molecular}.

Darwin’s theory of descent with modification requires that information be passed down from one generation to the next (heritability), that it can vary (through mutation in biology), and that there is a process of reproduction. Biological systems fulfill these conditions, but it has become clear that other systems do so as well. As such, phylogenetic techniques have been applied to language history and to cultural evolution \citep{Heggarty2006, o2025archaeology}. While the units and mechanisms differ, the basic conditions are fulfilled.

Galactic history may also fulfill the conditions for a phylogenetic approach \citep{Jofre2017}. The chemical composition of stars can be used as a proxy for heritability \citep{Freeman2002}, as these compositions are inherited from the parent cloud from which they are formed. Thus, sibling stars formed from the same source would share a similar chemical makeup. This provides such stars with a ``common ancestor" and a set of relationships analogous to a biological tree. In turn, after stars explode a new cloud is formed and the next generation of stars ``inherits" its chemical composition.

However, while this process creates conditions suitable for phylogenetic reconstruction, there are important differences from biological evolution. Change in organisms occurs through mutation and then the operation of natural selection. In galaxies, the change is the enrichment of the interstellar medium (ISM) that occurs through the constant stellar nucleosynthesis and stellar death, which in turn drive chemical evolution by altering the relative abundances of future stellar populations \citep{Burbridge1957, Tinsley1979, Matteucci2012}. Thus, although there is a different process driving change, this evolutionary case is suitable for the application of phylogenetic methods.

The idea of stellar siblings sharing a chemical composition has been a fundamental driver for the collection of large samples of stellar spectra, such as GALAH \citep{DeSilva2015,2025PASA...42...51B}. The objective for works drawing from these surveys would be to perform strong chemical tagging, that is, to find stars with identical chemical patterns in order to identify the building blocks of galaxies \citep{Freeman2002}. Numerous studies have assessed the power and limitation of strong chemical tagging in the disk \citep[e.g.,][and references therein]{2021A&A...654A.151C, 2024A&A...688A.165S, 2025A&A...702A.267S}, finding that the chemical compositions measured in stars still do not contain all the information to fully reconstruct the open clusters with high accuracy. The reason for this, beyond the uncertainties in the abundance measurements, is the chemical mixing in the ISM, which leads to degeneracies in the stellar abundances \citep[e.g.,][]{Schonrich17, Esteban22}. Weak chemical tagging uses more information than just chemistry (e.g., ages or kinematics), and it has been explored as a general alternative to grouping stars according to their galactic components \citep{2025ApJ...985..129B}. Ages and kinematics can break the degeneracies mentioned above \citep{ness2019galactic}. However, weak chemical tagging is focused on any trait, regardless of its heritability. 

In contrast to chemical tagging, which aims to group stars into distinct groups, stellar phylogeny aims to build trees that allow visualization of how the groups might be connected to each other. So far only a few works have been published on stellar phylogeny, with some based on observational chemical data \citep[e.g.,][]{Jofre2017, Jackson2021, Walsen2024, jofre25}. While these results allow plausible reconstructions of the histories of the stars used, these histories are difficult to validate due to a lack of stellar phylogenetic models. \citet[][]{deBrito2024} and \citet[][]{2026arXiv260411974T} analyzed numerical simulations of a galaxy that evolved in isolation to provide model expectations. While observational data should provide the ‘‘true’’ history of any population of stars, simulations offer crucial insight on the probability that a reconstructed history based on observed data is accurate, given the assumptions about the chemical processes involved. This allows for greater control over the interpretation of the phylogenetic trees and a test of the method as a means of reconstructing galactic history. 

In order to apply these hypotheses to real galaxies, we must consider that galaxies do not live in isolation. In fact, our own galaxy has been assembled through continuous accretion events of galaxies that have different properties and thus different chemical evolutionary histories \citep{2004AJ....128.1177V, 2019ARA&A..57..375S}. The Milky Way halo is the best place to find such merger remnants since star formation ceased far enough in the past that a significant number of halo stars observed today come from such accreted galaxies. Identifying the progenitor galaxies and studying their properties is an active field of research \citep{2020ARA&A..58..205H, 2023MNRAS.520.5671H}, and it has motivated the new generation of spectroscopic surveys such as 4MOST \citep{2019Msngr.175....3D} and DESI \citep{2026OJAp....955260K}. 

Chemical tagging in the halo is still difficult because the building blocks are evolved galaxies. This implies that clustering algorithms in chemistry need to account for metallicity distributions \citep{Buckley2024}. Additionally, the faintness of halo stars makes the determination of detailed chemical abundances very expensive with current instruments. To identify the progenitor galaxies thus requires weak chemical tagging techniques involving the kinematics of the stars, but even then it is not clear how to define and select stars from the different progenitors \citep{2024MNRAS.527.2165C}. Clustering techniques including ages and kinematics for simulated data in the halo are actively discussed in the literature \citep{2025A&A...704A..40T, 2026MNRAS.547ag503S}. 

When tested on simulations, however, clustering in phase space and in chemistry only allows for the recovery of the most recent and largest accreted components, especially when the in situ population of the Milky Way is included in the clustering. 
Furthermore, applying these clustering techniques to observed data is not straightforward, as ages are not available for all halo stars, suggesting new methodologies for further grouping \citep[see also][for extensive discussion on clusters in data]{Buckley2024}. 

It is thus fundamental to ask how phylogenetic trees can be used to disentangle different chemical evolutionary histories in the cosmological context of galaxy evolution, assuming that the patterns of evolutionary change of different star formation histories lead to different branches in phylogenetic trees. The motivation for this work is thus to identify the conditions under which a phylogenetic tree built with the methods employed before in astrophysics \citep[in, e.g.,][and references therein]{jofre25} would separate different stellar populations experiencing different chemical enrichment histories into different branches. With this end in mind, we constructed a set of galactic chemical evolution models that provide stellar abundances as a function of time within an open-box, one-zone framework with \flexce\ \citep{andrews17}. In this experiment, we then constructed composite populations of models as a proxy for a dry merger. 

The paper is structured as follows: In Sect.~\ref{sect:methods} we give an overview of \flexce\ and phylogenetics, describing how the models were built and then used to construct the phylogenetic trees. In Sect.~\ref{sect:results} we show the results for the branch separation within our trees, considering the effects of error. We also analyze these trees in terms of the impact of each input parameter on the quality of separation. Additionally, we present specific models. Then, in Sect.~\ref{sect:discussion}, we discuss our results in the context of Galactic archaeology. Finally, in Sect.~\ref{sect:conclusions} we summarize our main results and offer suggestions for future work.

\section{Methods}\label{sect:methods}

\subsection{Chemical datasets with tt{flexCE}}\label{sect:flexce}

We used the code \flexce\ \citep{andrews17}, which calculates the time evolution of chemical abundances of various elements in a single zone.\footnote{One-zone implies that there is no spatial distribution of any quantity studied in these evolutionary models. } In short, the code considers a box of initial gas mass that forms stars at a constant star formation efficiency. Stars form following a specific initial mass function \citep[in this case of][]{Kroupa2002}. Timescales for metal production are based on the lifetimes of stars of different masses. The model tracks the production of elements in massive stars, which return to the gas reservoir via core-collapse supernovae (CCSNe), asymptotic giant branch (AGB) winds, and Type Ia supernovae (SNeIa). The yields from \cite{Limongi06}, \cite{Chieffi04}, \cite{karakas10}, and \cite{Iwamoto99} are used, depositing chemical elements into the gas reservoir considering the number of progenitor stars formed at a given episode. \texttt{flexCE} considers delayed enrichment (e.g., by assuming a SNIa delay-time distribution), but instantaneous and homogeneous mixing for the returned material. Inflow and outflow of gas is also allowed. At each time step of the simulation, the total mass of each chemical element in the gas reservoir is updated.

We ran models varying five different input parameters: initial gas mass ($M_0$, measured in solar masses), inflow mass reservoir ($M_{1}$, measured in solar masses), inflow timescale ($\tau_1$, measured in Gyr), outflow mass-loading parameter ($\eta$), and star formation efficiency ($\nu$, measured in yr$^{-1}$). We note that these are all quantities that are parametrized as input for \texttt{flexCE}. More specifically, the inflow of mass $M_{in}$ is calculated as

\begin{equation}\label{eq:inflow}
    \frac{dM_{in}}{dt} = \frac{M_1}{\tau_1}e^{-t/\tau_1},
\end{equation}

\noindent and the outflow of mass $M_{out}$ is calculated as 

\begin{equation}\label{eq:outflow}
    \frac{dM_{out}}{dt} = \eta \ \mathrm{SFR},
\end{equation}

\noindent where ``$\mathrm{SFR}$" is the star formation rate, which in this case follows the Schmidt star formation law \citep{scmidt59} considering a power law index $N=1.0$. This means that 

\begin{equation}\label{eq:sfr}
    \mathrm{SFR} = \nu \ M_{gas},
\end{equation}

\noindent where $M_{gas}$ is the total mass of gas in the reservoir at time $t$ available to form stars. More details about these input parameters and their assumptions can be found in \citep{andrews17}. 

In our specific case, we always evolved boxes for 12 Gyr and followed the abundances of C, N, O, Na, Mg, Al, Si, S, K, Ca, Ti, V, Cr, Mn, Fe, Co, and Ni, which are the default elements provided in \flexce. 
We saved 400 time steps, evenly distributed within the 12 Gyr time span. 
Chemical abundances will evolve as long as there is sufficient mass in the box reservoir. In our case all gas reservoirs are isolated, assuming the initial mass function of \cite{Kroupa2002}, an exponential SNIa delay time distribution (DTD) for the SNIa, an exponential mass inflow function, an inflow composition according to a Big Bang nucleosynthesis abundance pattern, an outflow such that the ambient ISM is ejected in the wind, and a cold ISM.

\subsection{Fiducial models}\label{sect:fiducial}
We considered two fiducial models: a galaxy with a final mass similar to the Milky Way (hereafter {\tt mw-fid}) and a galaxy with a final mass similar to a dwarf-like galaxy (hereafter {\tt dwarf-fid}). Specifically, we ran both models for 12 Gyr and  
based our input parameters on the values for the Milky Way and the Gaia-Sausage/Enceladus from \cite{Limberg2022}. For the \mwfid\ model, we considered $M_0 = 2\times 10^{11} \mathrm{M_{\odot}}$, $M_1 = 3.5\times 10^{11} \mathrm{M_{\odot}}$, $\tau_1 = 6 ~\mathrm{Gyr}$, $\eta = 2.5$ and $\nu = 1.5\times10^{-9}\mathrm{yr}^{-1}$. 
For the \dwfid\ model we considered  $M_0 = 0.3\times 10^{11} \mathrm{M_{\odot}}$, $M_1 = 0.6\times 10^{11} \mathrm{M_{\odot}}$, $\tau_1 = 2.5 ~\mathrm{Gyr}$, $\eta = 5$ and $\nu = 1\times10^{-9}\mathrm{yr}^{-1}$.

The \mwfid\ model has the same input parameters as the fiducial model studied in \cite{andrews17}, except for $\nu$, which in this case is slightly higher in order to match with the observations of \cite{Horta2021}. As for the \dwfid\ model, we chose to lower $\eta$ from 6—as set in \cite{Limberg2022}—to 5 in order to constrain the range of $\eta$ used throughout this work.

These models are not intended to replicate the true Milky Way or a true dwarf galaxy; by construction, they have a very simple evolutionary history, unlike real galaxies. Also, many dwarf galaxies in the real Universe do not experience extended chemical evolution over cosmic timescales, as they are cannibalized by another galaxy \citep{Brown2012, Belokurov2013} or become quenched through processes such as reionization and ram pressure stripping \citep{Simpson18, Boselli22}.

\begin{figure} 
   \centering
   \includegraphics[width=3.3 in]{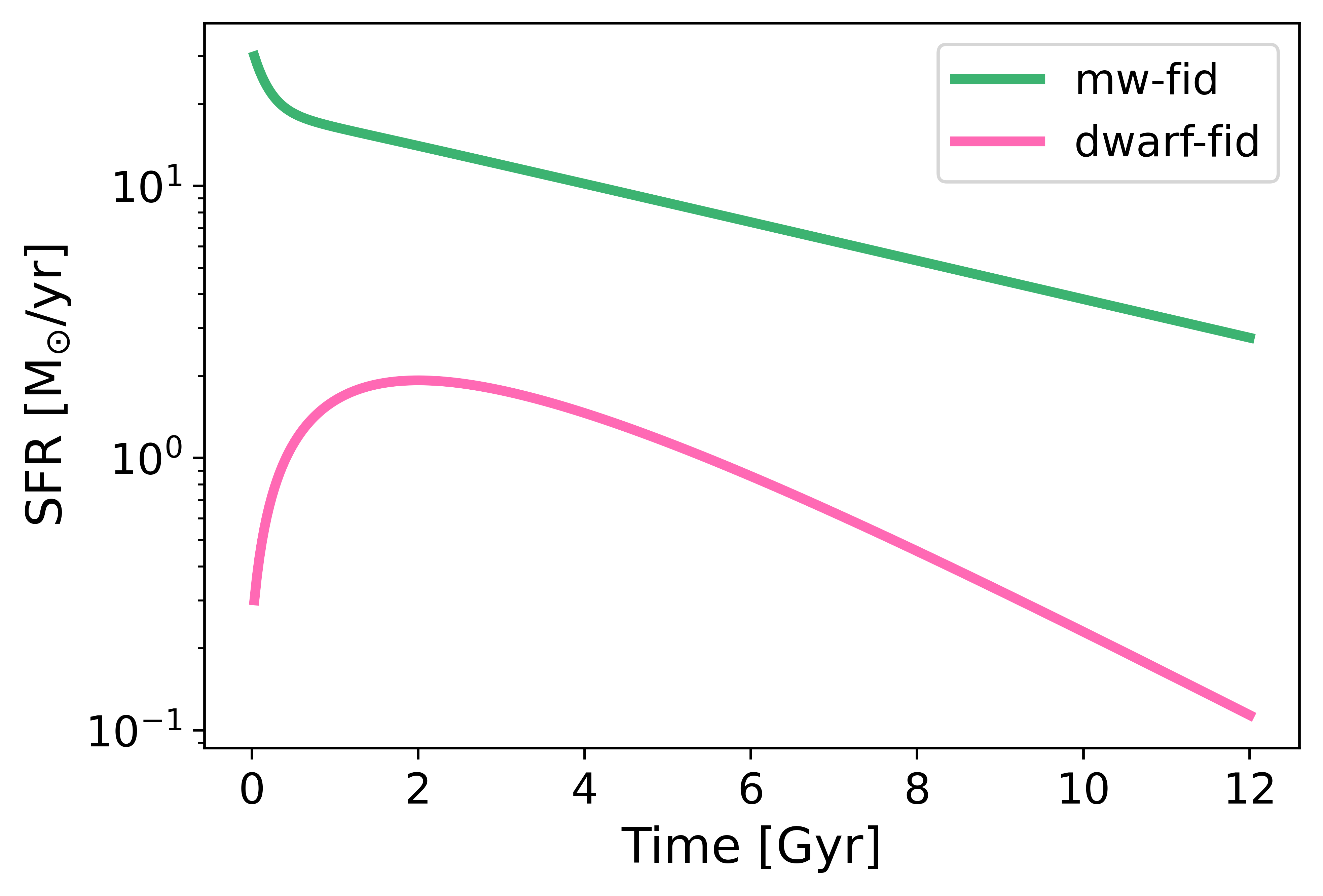}
        \caption{Star formation histories of our two fiducial models: \texttt{mw-fid} (in green) and \texttt{dwarf-fid} (in pink).}
 \label{fig:sfh mw dwarf}
\end{figure}

Figure~\ref{fig:sfh mw dwarf} shows the star formation history for both fiducial models. The SFR for each model was calculated following Eq.~\ref{eq:sfr}. One can see that the SFR of the \mwfid\ is higher than the one from the \dwfid\ at all times. Additionally, while the SFR of the \mwfid\ is at its highest at the first time step and steadily decreases over time, the \dwfid\ has a peak with a 1-2 Gyr delay. This is due to the way \flexce\ calculates the SFR, making it proportional to the gas mass within the system.  
We note that in \flexce, the effects of the outflow rate are not immediately evident \citep[][their Fig.~3d]{andrews17}. As the other types of stars die, their material is expelled from the system as well, leading to the sharper quenching of star formation seen in Fig.~\ref{fig:sfh mw dwarf}.

\begin{figure} 
   \centering
     \includegraphics[width=3.3 in]{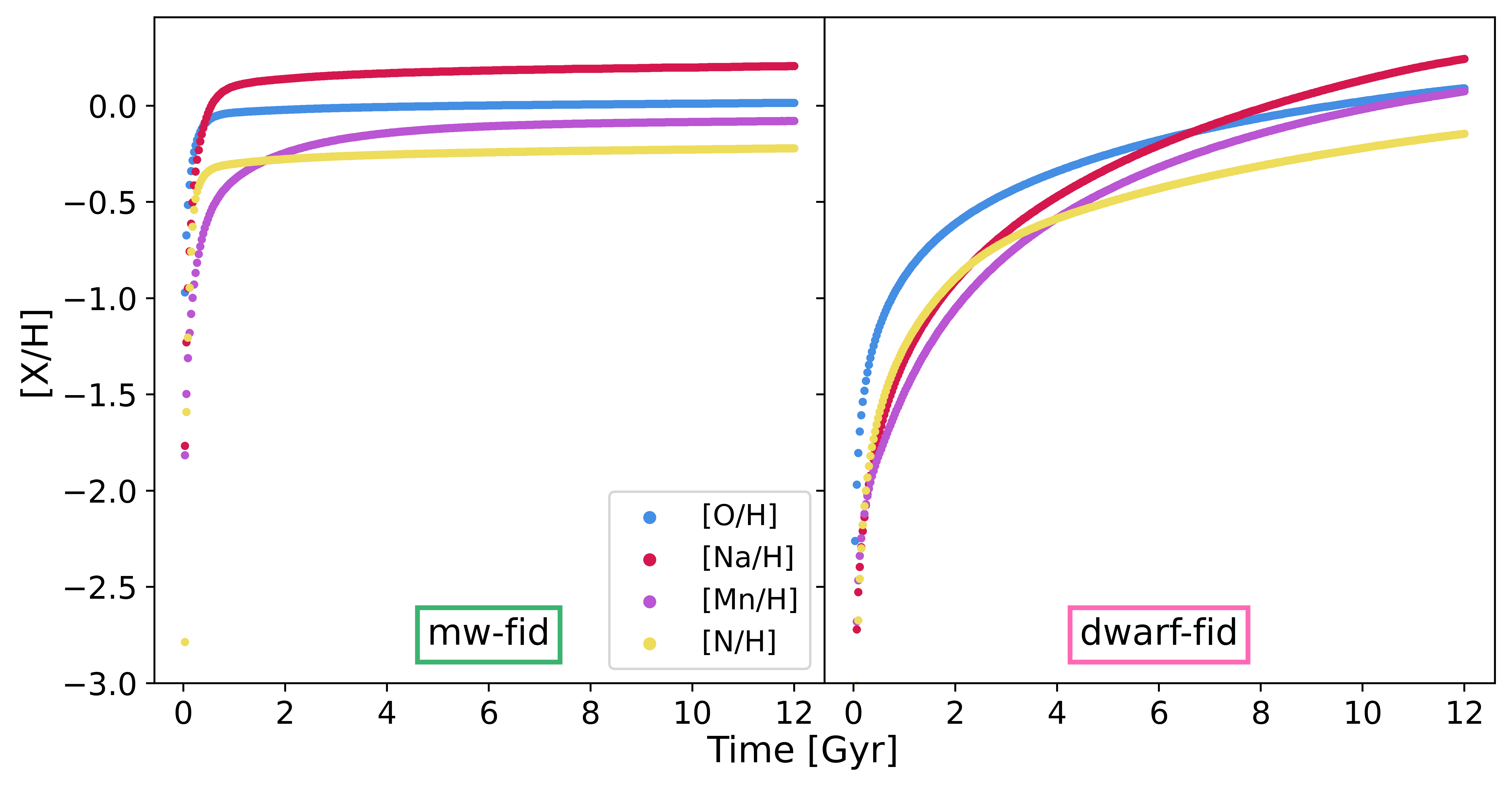}
     \caption{Chemical abundance ratios as a function of time for the \mwfid\ model (\textit{left}) and the \dwfid\ model (\textit{right}). We show the evolution of [O/H] in blue, [Na/H] in red, [Mn/H] in purple, and [N/H] in yellow. }
 \label{fig:abund_time}
\end{figure}

The difference in the star formation history between both of these models impacts the chemical enrichment, which can be seen in Fig.~\ref{fig:abund_time}. Here we plot the evolution of the abundance ratios of N, O, Na, and Mn over H in different colors. We choose these elements as representative of different nucleosynthetic channels. More specifically, oxygen is produced mainly by CCSNe \citep{Meyer2008}, and its production has a very weak dependency in [Fe/H]. O is thus more strongly linked to CCSNe, and is widely used for studying chemical evolution driven by massive stars—and in consequence, rapid star formation histories \citep{Tissera19, Smith2021, Chruslinska2024, Boardman2025}. Nitrogen, on the other hand, is produced through the CNO cycle in main sequence stars, where C and O get converted into N. After the various mixing mechanisms caused by dredge-up inside red giant branch stars, AGB stars can expel the new N into the ISM via winds \citep{Ritter2013, DiCriscienzo2016}. 

Sodium is an odd-Z element which is also synthesized in CCSNe. The difference between Na and O is the strong dependency in metallicity for its production; in other words, the more metal-rich the progenitor stars are, the more Na they produce. Similarly to Na, manganese has a strong dependency in metallicity for its production. Its main source is SNeIa, although some production comes from CCSNe, which becomes more evident at low metallicity. This has made these two element abundances important for disentangling different star formation histories, notably between stars formed in the Milky Way or in dwarf galaxies \citep{hawkins15, das20, buder22}. Other elements might behave as these four, or might be a mix between them. Extensive discussions of their production sites and how they impact the evolution in \flexce\ can be found in \citet[][their Sect.~4]{andrews17}.

In Fig.~\ref{fig:abund_time} we plot the evolution of the four elements discussed above. 
One can see how for the \mwfid\ model, the elements increase sharply during the first few gigayears, which is when the SFR is the highest. On the other hand, the increase of these elemental ratios for the \dwfid\ model is more gradual, spanning the entire history of the simulation.  
We can understand why [Mn/H] takes longer to reach its highest value, as SNeIa have a delay compared to CCSN. We furthermore see that [Na/H] quickly reaches higher values than [O/H], which might be caused by the metallicity dependency, which also increases quickly in this model. Indeed, the Na yield of \cite{Chieffi04} is greater than the O yield for stars above $20~M_\odot$ at their highest metallicity.
The evolution of the elements in the \dwfid\ model shows that [O/H] increases steadily, but its largest production happens early on. [Na/H], on the other hand, becomes higher than [O/H] much later than for the \mwfid\ model. This is because the overall chemical enrichment is slower, due to the lower SFR, and so the Na production is lower if there is lower metallicity. 
We note that even the parameters of the fiducial model adopted in \citep{andrews17} shows a significant increase of [Na/H] over time. The purpose of our experiment is to compare models. Since these parameters are adopted for all models, the impact on the effect of Na is similar and the relative differences remain unaffected. Abundance planes as a function of Fe for the fiducial models can be further found in Appendix~\ref{app:planes}.

\subsection{Phylogenetic analysis and null hypothesis}

A phylogenetic tree is a graph that serves as a visualization tool to display the similarities of objects in a dataset, in order to study the hierarchical relationships between them. If the traits trace heritable information, then the phylogenetic tree can be used to trace the evolution of these traits within the group \citep{baum2005tree}. This can be done by looking at the branching pattern and the relationships displayed between the internal nodes and the external nodes or tips. In our case, the heritable traits are the chemical abundances. In the \flexce\ models, we use the chemical composition of the ISM at the time of the star formation to represent the composition of the stars formed.

\begin{figure} 
  \centering
   \includegraphics[width = 3.3 in]{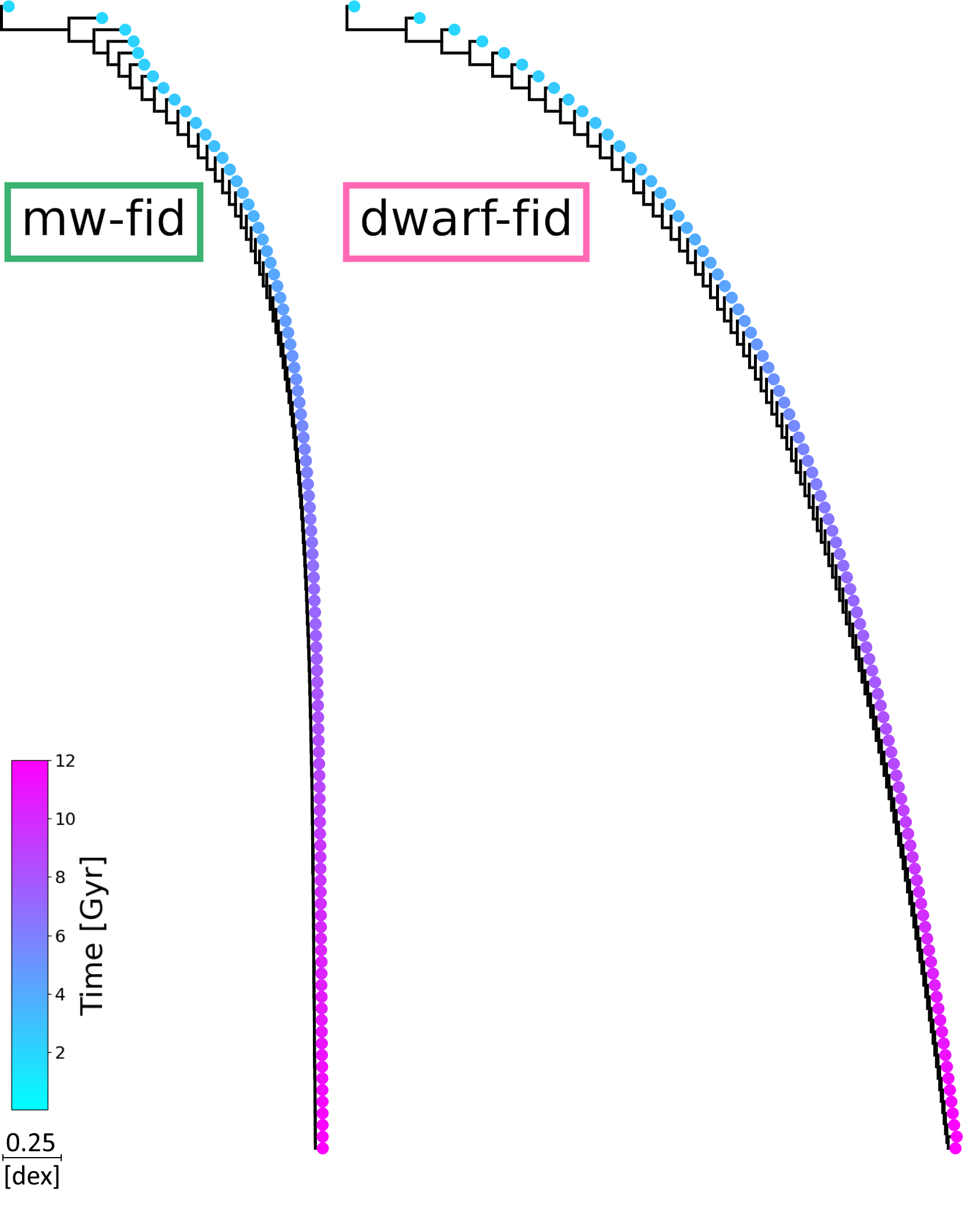}
   \caption{NJ phylogenetic trees of the \mwfid\ model (\textit{left}) and the \dwfid\ model (\textit{right}). Tips are colored by their age, and the scale bar is in dex.}
   \label{fig:trees}
\end{figure}

We plot the phylogenetic tree for the \mwfid\ (left) and the \dwfid\ (right) models in Fig.~\ref{fig:trees}. In these plots, we consider [X/H] abundance ratios, where X are all the elements followed in the model as introduced in Sect.~\ref{sect:flexce}. Each tip is the chemistry of a \flexce\ time step, which is color-coded by its age. We are considering the abundances and their evolution as displayed in Fig.~\ref{fig:abund_time}. Trees were built following the methodology of \cite{deBrito2024, Walsen2024, jofre25, 2026arXiv260411974T}; namely, computing a pairwise distance matrix, in which each element of the matrix is the Euclidean distance of all [X/H] abundance ratios of a given time step of the \flexce\ model. We note, however, that trees constructed using the distance method can employ other distance metrics, such as Manhattan \citep{Jofre2017, Jackson2021}. It is also possible to perform a phylogenetic analysis using other measurements of chemical elements, such as [X/Fe] or even chemical clocks \citep[see][for a discussion of tree shapes using different chemical element ratios]{Jackson2021}. Our current choice of using Euclidean distances and the abundance ratios in [X/H] is to better interpret and compare our results with the latest works using numerical simulations \citep{deBrito2024, 2026arXiv260411974T}. Both here and in the cited works, using [X/H] allows for phylogenetic interpretation when abundance ratios have a monotonic chemical enrichment over time (see Fig.~\ref{fig:abund_time}). This might not happen in galaxies which experience wet mergers, because they can gain substantial H and so drop their [X/H], but that is not the case here. The impact of the phylogenetic analysis using different distance metrics or traits is active topic of research \citep{bromham2016introduction} and thus is subject for a different study.  

From the distance matrix, the phylogenetic trees were built using the neighbor-joining method \citep[NJ,][]{Saitou1987}. The total chemical difference between the time steps is reflected in the branch lengths of these phylogenetic trees. Further details on the algorithms used to construct trees can be found in Appendix~\ref{app:trees}.
In this case, the branch length is measured in dex from the abundances. It is worth commenting that the trees in this work are plotted in rectangular form, and as such only the horizontal branches represent the chemical distance measured in dex. Vertical branches have arbitrary units and have no astrophysical representation. In Fig.~\ref{fig:trees} we can see how the branch lengths become shorter along the trees.

The \flexce\ models are one-zone, meaning that the chemical elements generated in one star forming episode are fully incorporated and processed into the next episode. This, along with the fixed input parameters, results in all [X/H] abundance ratios increase monotonically (see also Fig.~\ref{fig:abund_time}). Given that our phylogenetic trees are built solely on the chemical distances of the same model at different times, this evolution translates into a ``caterpillar" tree \citep{jofre25}. The imbalance in these phylogenetic trees has been previously noted as a characteristic feature in galactic chemical evolution \citep{2026arXiv260411974T}. Caterpillar trees can be attributed to anagenesis\footnote{The gradual evolution of a single lineage through time.} \citep[see][for more discussions]{Jackson2021}. 

Even though in both cases the phylogenetic trees have a caterpillar shape, the \mwfid\ phylogenetic tree is shorter than the \dwfid\ one. This is related to the star formation histories of both models, which are different (see Fig.~\ref{fig:sfh mw dwarf}). Even though more stars are made in the \mwfid\ model than the \dwfid\ model, the chemical enrichment in the \mwfid\ model happens very quickly in the first 2 Gyr and stays more or less constant thereafter (see Fig.~\ref{fig:abund_time}). The \dwfid\ model, on the contrary, continuously enriches its gas with more chemical elements over time.  This effect implies a larger difference in chemical abundance between time steps for the \dwfid\ model than for the \mwfid\ model, which translates into distance matrices with larger values, and thus larger branch lengths. It is worth stressing that this does not mean that the \dwfid\ model achieves higher absolute abundances, the tree is showing the relative difference, not the total abundances. The \dwfid\ model continuously changes its [X/H] with larger amounts than the \mwfid\ model, especially at later times. 

Discussion on these differences can be framed through the work of \cite{deBrito2024}, who compared phylogenetic tree lengths calculated from different regions of a simulated disk galaxy in isolation. There, longer trees were attributed to a higher chemical enrichment caused by a higher SFR. The overall star formation history in that simulation had the same shape, but a different scale across the disk. In contrast, here the star formation histories are different. Therefore, attributing a longer tree to a higher SFR is not correct if the star formation histories are different.

\begin{table}
    \centering
    \caption{Grid used for generating \flexce\ models.}
    \begin{tabular}{|c|cccc|c c |c|}
        \hline
        Input & A & B & C & D & MW&DW& Unit \\
        \hline
        \hline
         $M_0$ & \rule{0pt}{1.2em}0.5 & 2.0 & 3.5 & 5.0 & 2& 0.3&$10^{10} M_\odot$ \\
          $M_1$ & 0.9 & 3 & 6 & 9 & 3.5&0.6&$10^{11} M_\odot$ \\
          $\tau_1$ & 1 & 4 & 7 & 10 & 6&2.5&Gyr \\ 
          $\eta$ & 0.5 & 2 & 3.5 & 5 & 2.5& 5 & \\ 
          $\nu$ & 0.6 & 1.2 & 1.8 & 2.4 & 1.5&1&$10^{-9} \mathrm{yr}^{-1}$ \\ [1ex]
        \hline
    \end{tabular}
    \label{tab:grid}
    \tablefoot{For a given model each of our input parameters can take one of four predefined values, labeled as A, B, C, or D and increasing in that order. The input parameters of the fiducial models are listed under the columns MW and DW.}
\end{table}

\subsection{Mimicking dry mergers of two systems with \texttt{flexCE}}

Given our hypothesis that the phylogenetic trees of Fig.~\ref{fig:trees} reflect the history of a self-evolving system in isolation, and that the phylogenetic patterns are different from each other if the histories are different, we are interested in knowing if there are two models in \flexce\ that lead to a tree which divides the models into two different branches.  In other words, we are interested in studying the prospects of phylogenetics to disentangle stars that are a fossil record of different evolutionary histories. Such a combination of stars from different origins can be found in the halo \citep{2004AJ....128.1177V, 2020ARA&A..58..205H}. 

To answer this question, we generate \flexce\ models (in isolation) with different input parameters as explained above.  The input parameter values we adopt are indicated in Table~\ref{tab:grid}, where we also list the input parameters of the fiducial models for reference. With these outputs, we combine two such models and build a single tree. To build one tree, we calculate a pairwise distance matrix from a table which includes the chemical abundances of both models concatenated and run the NJ algorithm on these matrices. Our goal is to find which combination of parameters gives us trees with two branches, in which each branch contains the data of one of the models. This is relevant because, as recently discussed by \citep{2026arXiv260411974T}, stellar populations which have evolved in different environments can mix in phylogenetic trees.  

In this study, we combine all the models with each of the fiducial models. This allowed us to study and compare similar models and very distinct models. It is important to emphasize that in this experiment, each population was allowed to evolve independently from the other, meaning that even in the combined models there is no intermixing of the ISM between either of the progenitor models. We stress that our experiment is simplistic, combining only two models, whereas a place like the Galactic halo hosts many more accreted galaxies \citep[see also][]{2025arXiv251207780G}. Also, our models evolve for 12 Gyr, which is not the case for most dwarf galaxies. Finally, our combination considers approximately the same number of data points per model (Sect.~\ref{sect:results}), but in a real dataset the percentage of stars from a specific origin is not known a priori. With all these considerations, this experiment is the first one of its kind, and so it offers a first hint on how to interpret real data with phylogenetics. Therefore, the experiment needs to be kept simple. 

\begin{figure} 
\centering
\includegraphics[scale=0.6]{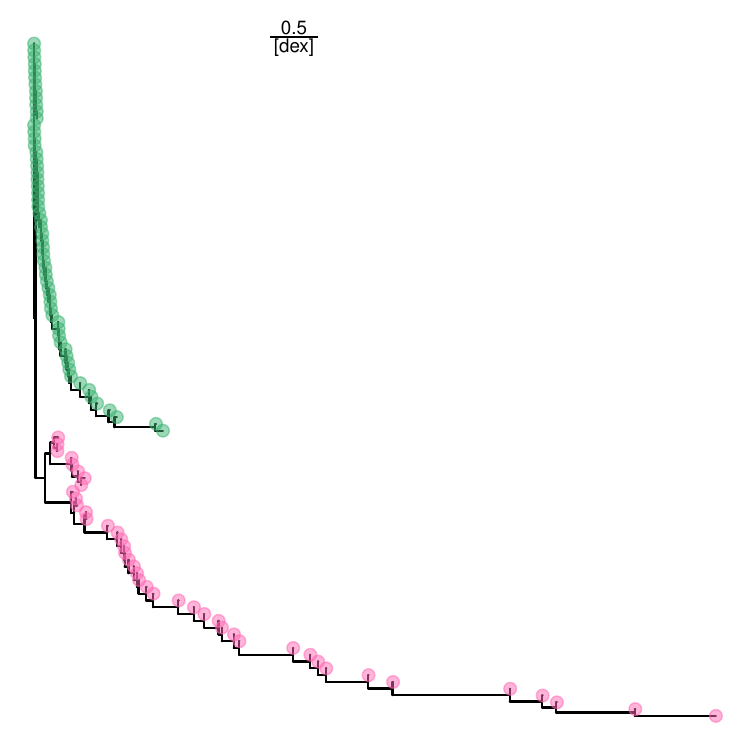}
       \caption{Phylogenetic tree of a random selection of 100 data points from \mwfid\ and \dwfid\ models together. As before, green represents the \mwfid\ model and pink the \dwfid\ model. Horizontal branch lengths correspond to the pairwise Euclidean distance of all chemical abundances. The scale lengths of such distances are indicated in the top-middle bar.  }
   \label{fig:tree_mixed}
\end{figure}

In Fig.~\ref{fig:tree_mixed} we plot the NJ tree built from the combined data of the \mwfid\ and the \dwfid\ models, selecting 100 data points randomly as done in the subsequent steps of this analysis (see further discussions in Sect.~\ref{sect:results}). 
The tree subsequently helps to understand how the dataset is distributed in chemical space. The tips are colored by their origin, namely the \mwfid\ and \dwfid\ models. In this case, we can see that the phylogenetic tree has two branches, and that each branch contains the data of a different model. Therefore, for this tree, we can say that the NJ algorithm is able to separate the \mwfid\ from the \dwfid\ model into two separate branches (histories), despite the overlap of some of the abundance ratios in their chemical trends discussed in Appendix~\ref{app:planes}.

Both branches are connected through the most metal-rich data points of each model. In  Figs.~\ref{fig:abund_time} and \ref{fig:alpha_odd_z}, we can see that the models tend to become more similar to each other with time (or metallicity). This is because the SFRs of these models are higher and few CCSNe significantly impact the pristine gas, so models are most distinct at the start, when metallicity is low. The NJ algorithm thus places these data points farther apart than when the abundances converge into similar enrichment levels. The final enriched chemical pattern, however, is different for both models, which is why the NJ algorithm still shows a larger branch between them than between adjacent nodes of the same model (see more discussions in Sect.~\ref{sect:ga}).  

For our experiment, we have 1024 grid models to combine with the fiducial models. All models, including the fiducial ones, can be accessed through Zenodo. 
Each grid model has different input parameters which cover a five-dimensional grid of initial values listed in Table~\ref{tab:grid}. Finding the combination of parameters that results in a tree with two clean branches requires an automatic search in the phylogenetic space, as visual inspection is not feasible. Therefore, given a phylogenetic tree divided into two branches, we must first assess when each branch contains tips from only one model.

To do so, we applied a heuristic algorithm to detect the split that yields the most balanced tree. In phylogenetics, any internal edge induces a split or bipartition that divides the dataset into two disjoint groups \citep{felsenstein1985phylogenies}. Because our one-zone models produce monotonic chemical enrichment—resulting in highly unbalanced, caterpillar-like topologies, we specifically search for the bipartition that minimizes structural imbalance. We are interested in finding the bipartition  that divides the tree into two branches with the most similar number of nodes possible. To this end, we count every internal bipartition of the tree, pick the one that divides the data most evenly by count, and then measure how well that cut separates the two models.
In this case, we refer to them as ``grid" and ``fid."  We define a purity $P$ value which corresponds to the average of the percentage of ``grid" and ``fid" tips in each branch:

\begin{equation}\label{eq:purity}
    P = \frac{p_\textrm{fid} + p_\textrm{grid}}{2},
\end{equation}

\noindent where $p_\textrm{fid}$ is the percentage of nodes from the ``fid" model present in the ``fid"-dominated branch, and analogously for $p_\textrm{grid}$.

A tree separates well into two branches with two models if $P>0.9$. Although this cut might seem arbitrary, our results do not significantly change if we adopt a cut of  $P=0.8$ (see Appendix~\ref{app:p08}). It is relevant to stress that this method works well for two models which contribute to roughly the same number of tips in the dataset, which is the starting point of this experiment. 

\begin{figure}[t] 
\centering
\includegraphics[scale=0.6]{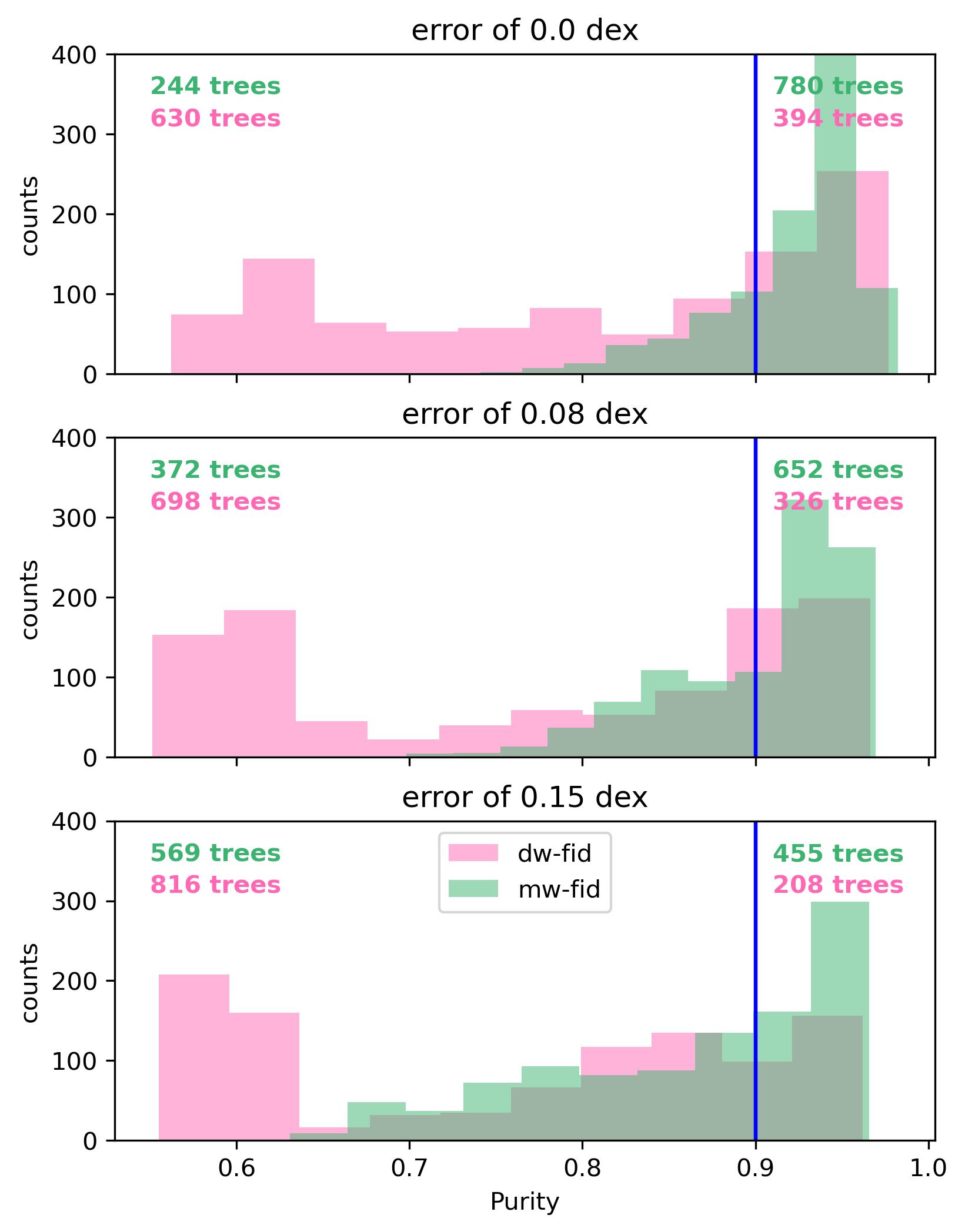}
       \caption{Distribution of purity, $P$, for the combined grid and fid models, whose abundances have no error, an error of 0.08 dex, and an error of 0.15 dex.  The green (pink) histogram indicates the grid models combined with \mwfid\ (\dwfid). The vertical blue line indicates the $P>0.9$ cut defined as good separation between models within the tree. The number of trees for each fid model above and below the cut are indicated in each panel with the corresponding color. }
   \label{fig:purities}
\end{figure}

\section{Results}\label{sect:results}

\subsection{Random tips with errors}
Combined models have 800 data points with chemical abundances (400 for the ``fid" model and 400 for the ``grid" model). We evaluate the impact on purity when taking random samples of 100 data points with different levels of uncertainties in the abundances.
In order to account for the impact of uncertainties, 
we associated each data point with a Gaussian centered at the abundances of said point, with a FWHM that mimics the precision of abundance measurements in observations. The perturbed abundance was then obtained by randomly sampling a value from this Gaussian.
We chose 100 tips because this is approximately the number of stars used in observational studies that enable us to draw conclusions about different star formation histories \citep{deBrito2024, jofre25}, though tests with a higher number of tips showed no significant impact on our conclusions (see Appendix \ref{app:rf_200}).  
The distribution of the purities for all grid models, compared to both fiducial models, can be found in Fig. \ref{fig:purities}. The vertical blue line indicates our threshold of good and bad separation by indicating $P=0.9$.

As can be seen, most trees in the \mwfid\ case separate cleanly when the abundances have no uncertainty, while in the \dwfid\ case there is a more even split between trees that separate and trees that do not. 
When adding uncertainties, we find that the confidence of separation decreases for the \mwfid\ case, whereas the purities in the \dwfid\ case remain relatively unaffected. From Fig.~\ref{fig:purities}, we find that at a conservative error of 0.15~dex, the number of trees for the \mwfid\ case that become poorly separated increases considerably. For that same error, the \dwfid\ has more trees that separate poorly compared to those that separate well, but the numbers do not vary as much between error values.

\subsection{Mapping input parameters with purity}\label{sect:mapping}

We used random forests to perform a classification algorithm that distinguishes between phylogenetic trees that are well separated and those that are not. This was done separately for the \mwfid\ model and for the \dwfid\ model. The algorithms work on the phylogenetic trees of combined models, where the input features correspond to the values of Table~\ref{tab:grid} and the output is a binary array of zeros and ones. Zero means no separation ($P<0.9$), while one means separation ($P>0.9$). 

Our random forest classification algorithm used the Python library {\tt sklearn} \citep{scikit-learn}. We created a training set by sampling randomly from 80\% of our models, and tested on the remaining 20\%, ensuring for each model a minimum accuracy of 90\% using a k-fold (k=10) cross validation.
Details of the setup of the classification and other outputs can be found in Appendix~\ref{app:rf}. 

\begin{figure}[t] 
\centering
\includegraphics[scale=0.45]{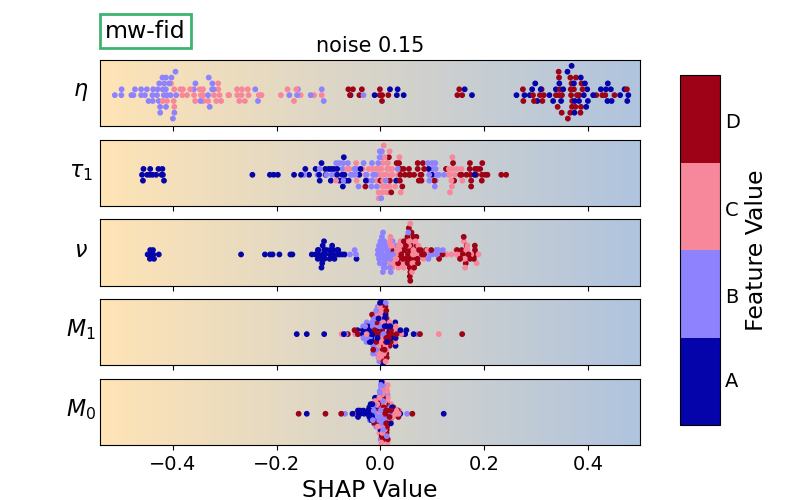}
\includegraphics[scale=0.45]{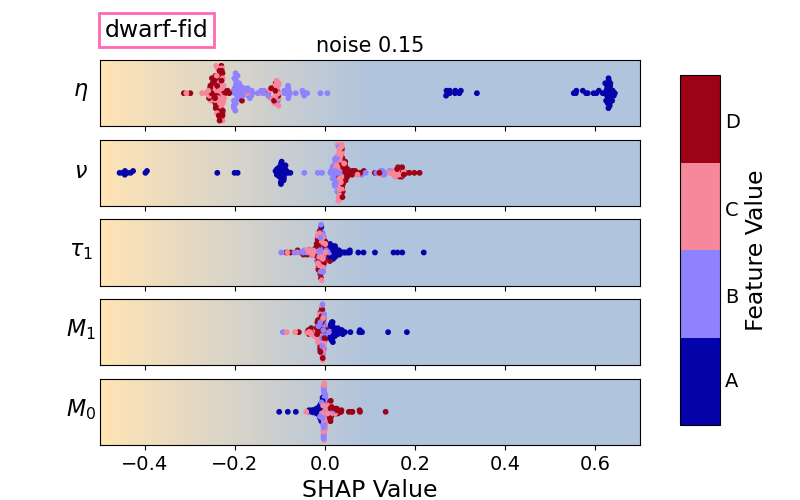}
       \caption{Importance of the input features for a random forest interpreted with Shapley values, which studies the condition $P>0.9$ for the models compared to the \mwfid\ (\textit{top}) and \dwfid\ (\textit{bottom}). }
   \label{fig:shap}
\end{figure}

The Shapley values of the features used for the training are plotted in Fig.~\ref{fig:shap} in the top panel for the \mwfid\ case and the bottom panel for the \dwfid\ case. The classification works as follows: If the sum of the Shapley values of a combination of five features is positive, then the model predicts a class 1 ($P>0.9$). If the sum is negative, then the model predicts a class 0 ($P<0.9$). This way of visualizing the random forest helps us find which feature has the larger impact in obtaining a given class.

Figure~\ref{fig:shap} thus sorts the features by importance in the classification. In both panels, we see that $\eta$ is the most influential input parameter when it comes to determining the purity. Then, $\nu$ is important at a second level for both models, but for the \mwfid\ case $\tau_1$ also plays an important role, while for the \dwfid\ case this feature is the least important. In both cases we see that $M_0$ and $M_1$ have little impact in the classification of the model, and therefore minimal impact on the topology of the phylogenetic trees.  

More specifically, it is possible to identify that the \mwfid\ model separates from models which have very low or very high values of $\eta$. This would make physical sense, since these are the input values which differ the most from the $\eta$ value adopted by the \mwfid\ model. Similarly, the random forest favors separating the \dwfid\ case for low $\eta$ values, which have the largest difference with the $\eta=5$ adopted for the \dwfid\ model.  

Regarding $\nu$, we see that for both models low $\nu$ tends to mix models, while other $\nu$ values have no significant impact in the separation of trees. For both fiducial models $\nu$ is relatively low; therefore, if similar $\nu$ values are adopted then trees do not separate. However, when the adopted values of $\nu$ differ more, the trees do not necessarily separate. This might be due to the dependency of other input features that also play a secondary role. A similar effect is seen for $\tau_1$ for the \mwfid\ case.

\begin{figure*}[t]
\centering
\includegraphics[scale=0.35]{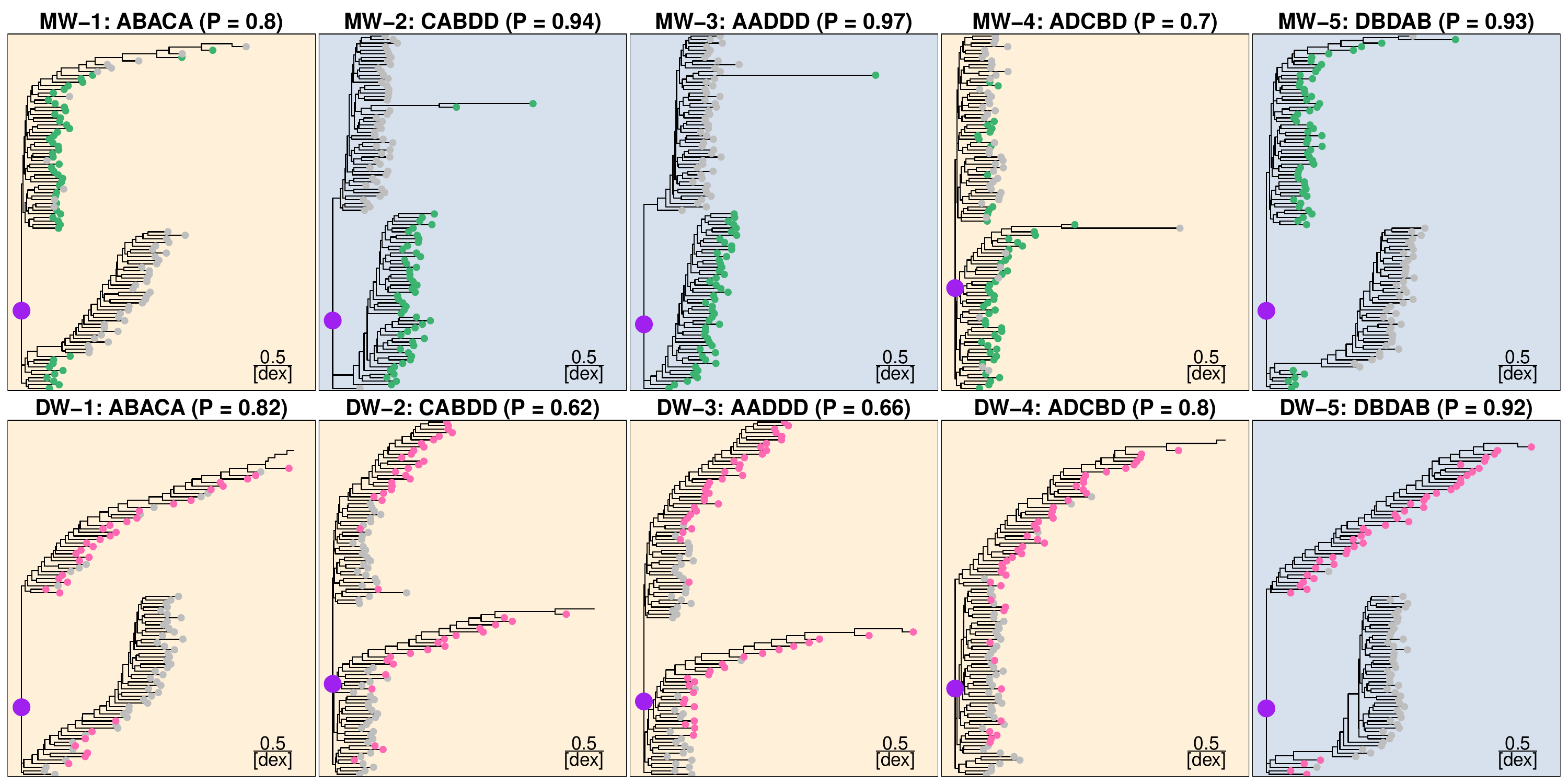}
       \caption{Examples of trees of \mwfid\ (\textit{top}) and \dwfid\ (\textit{bottom}) models considering uncertainties of 0.15 dex. Grid models have gray tips, while fid models have green tips for \mwfid\ and pink tips for \dwfid. Purity is indicated, and the background is yellow for $P<0.9$  and blue for $P>0.9$.} 
   \label{fig:examples_015noise_100}
\end{figure*}

\begin{figure*}[t]
    \centering
\includegraphics[width=0.192\linewidth]{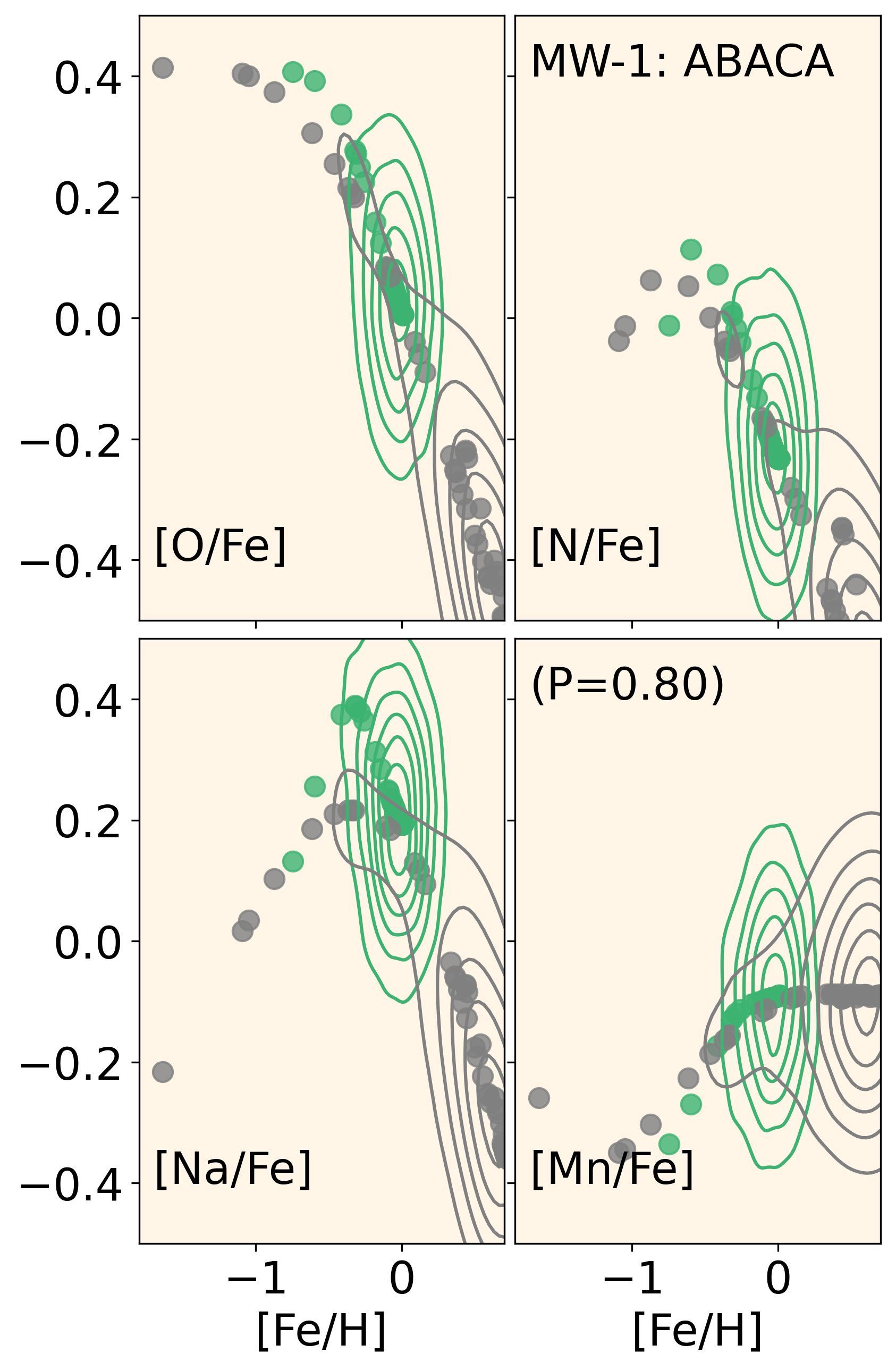}
\includegraphics[width=0.192\linewidth]{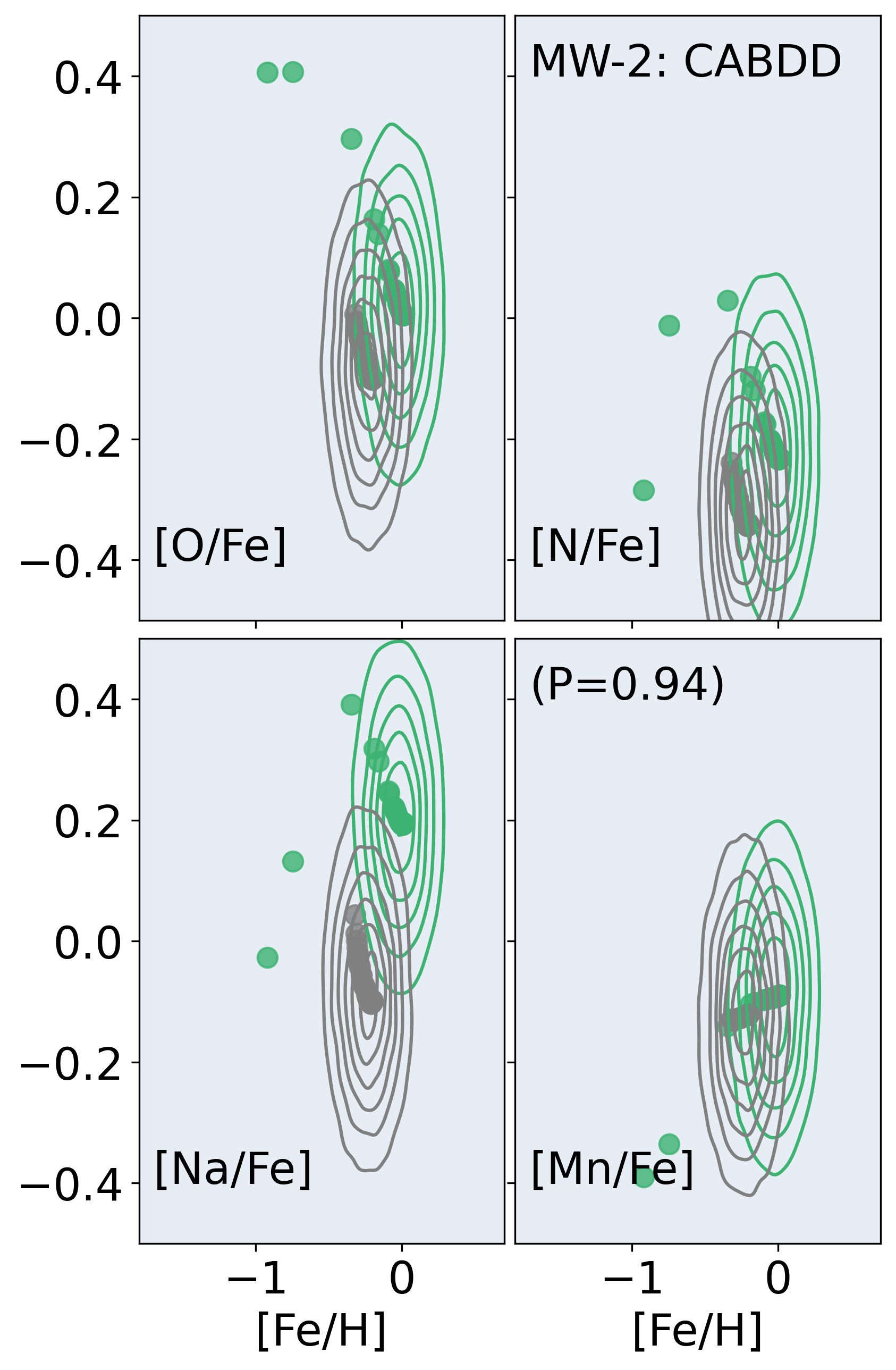}
\includegraphics[width=0.192\linewidth]{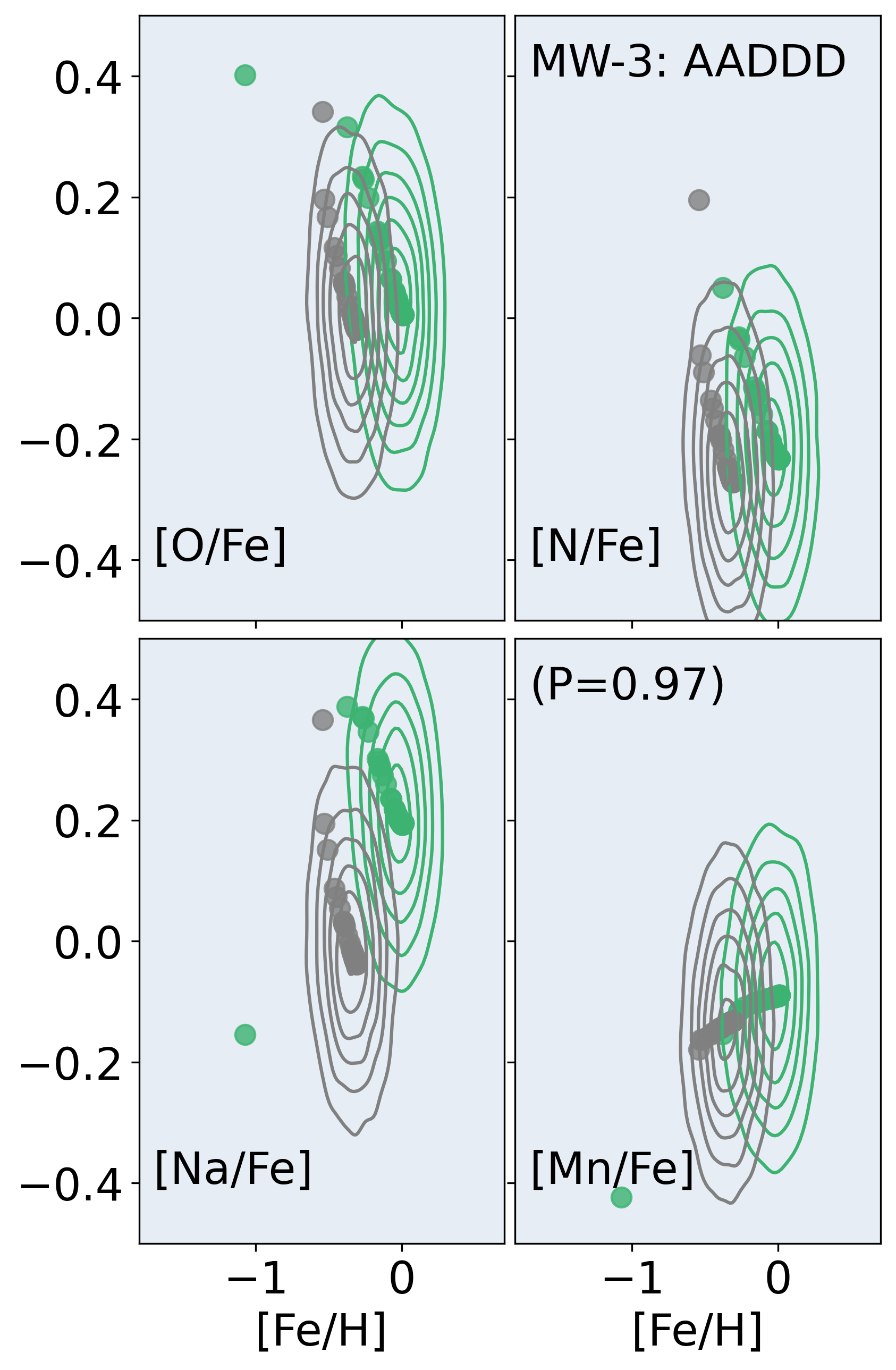}
\includegraphics[width=0.192\linewidth]{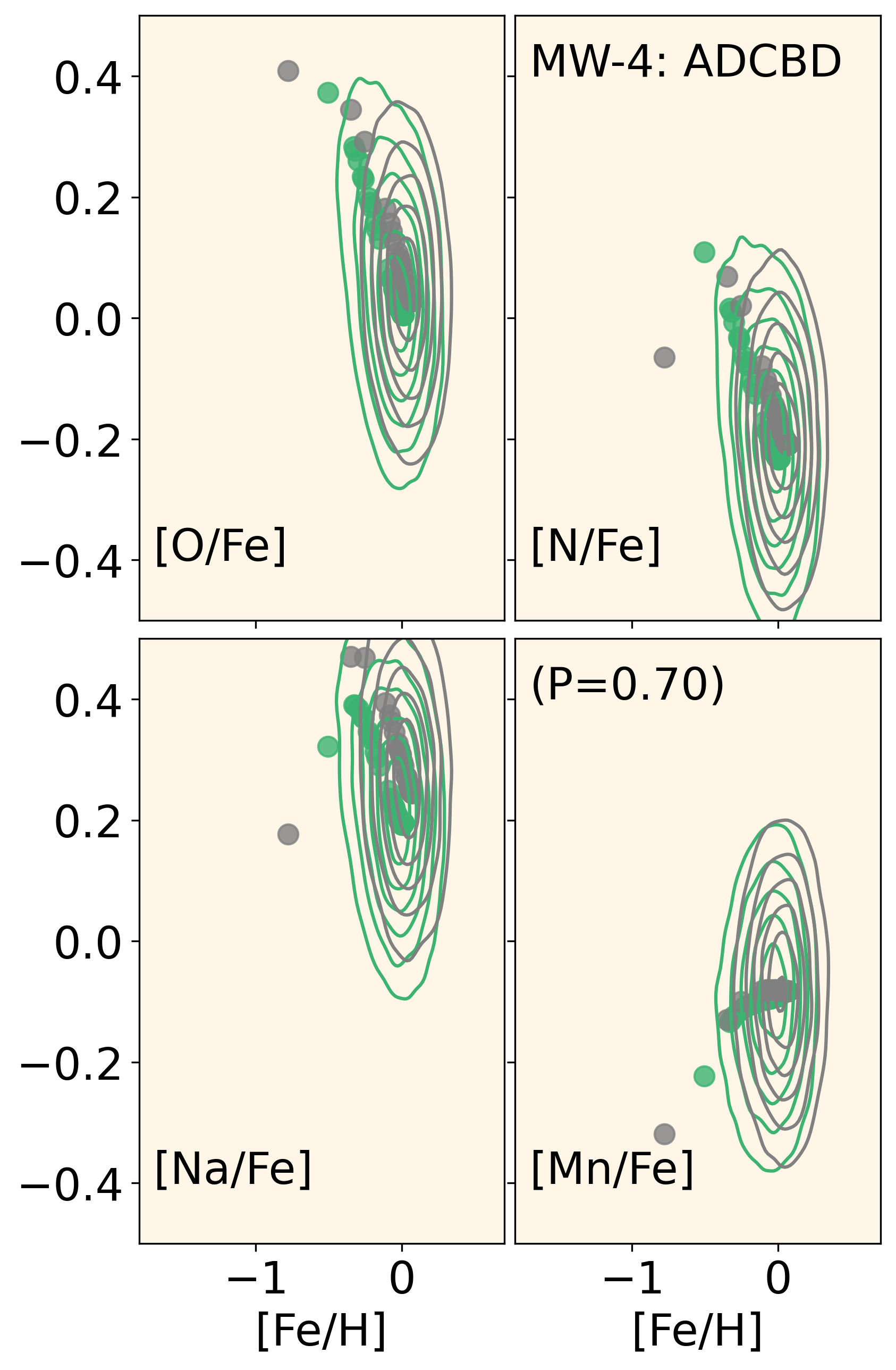}
\includegraphics[width=0.192\linewidth]{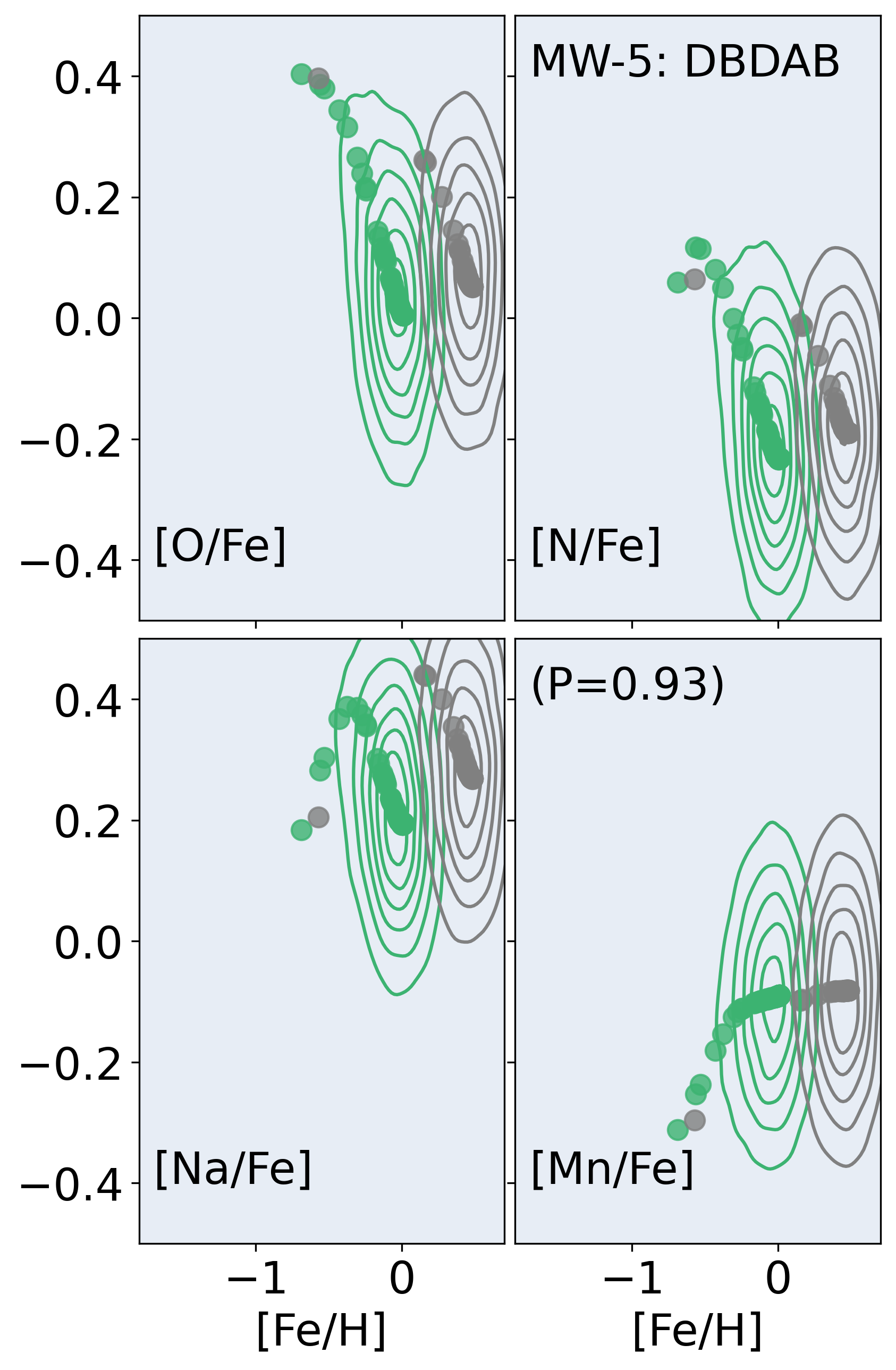}

\includegraphics[width=0.192\linewidth]{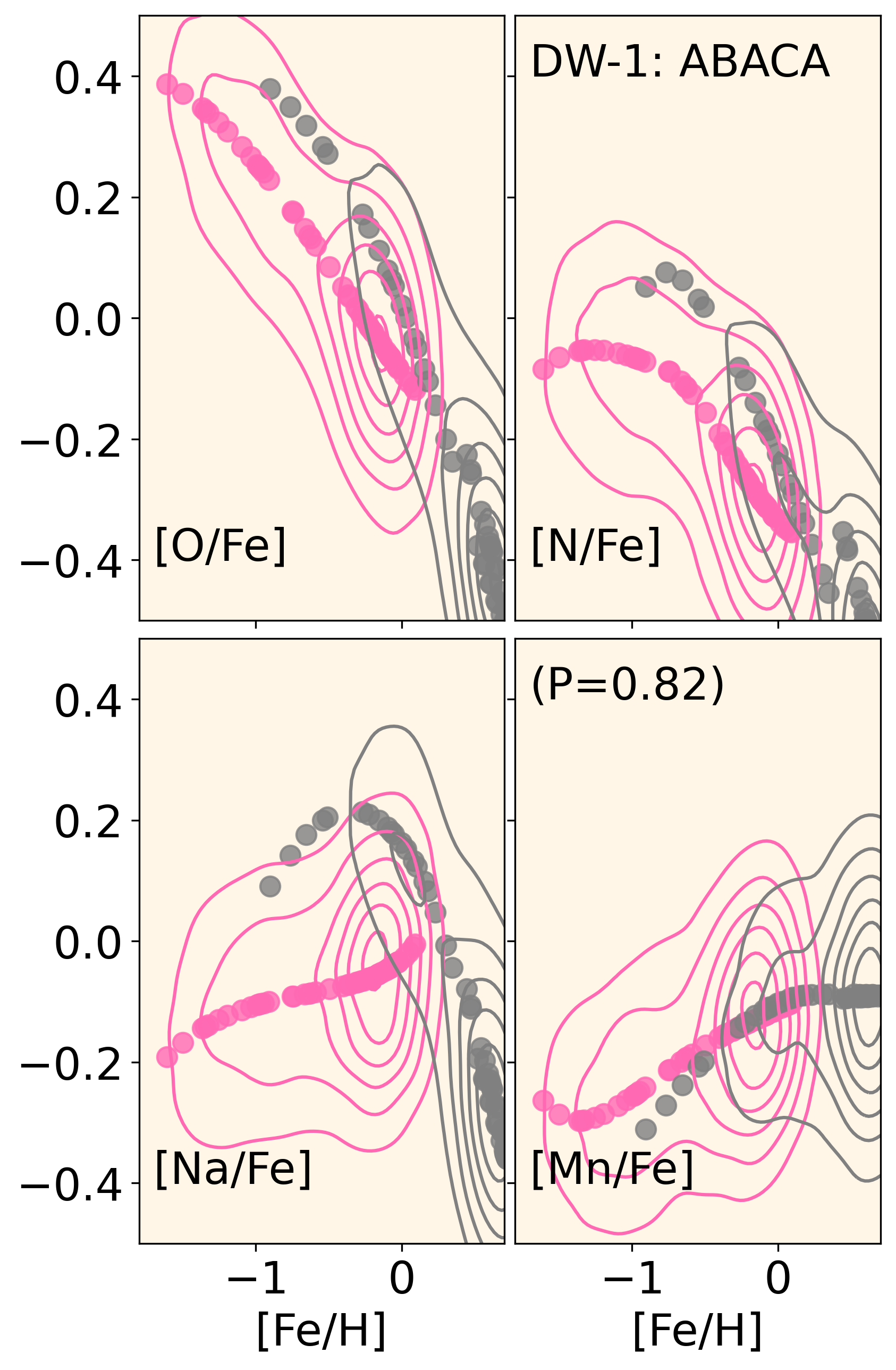}
\includegraphics[width=0.192\linewidth]{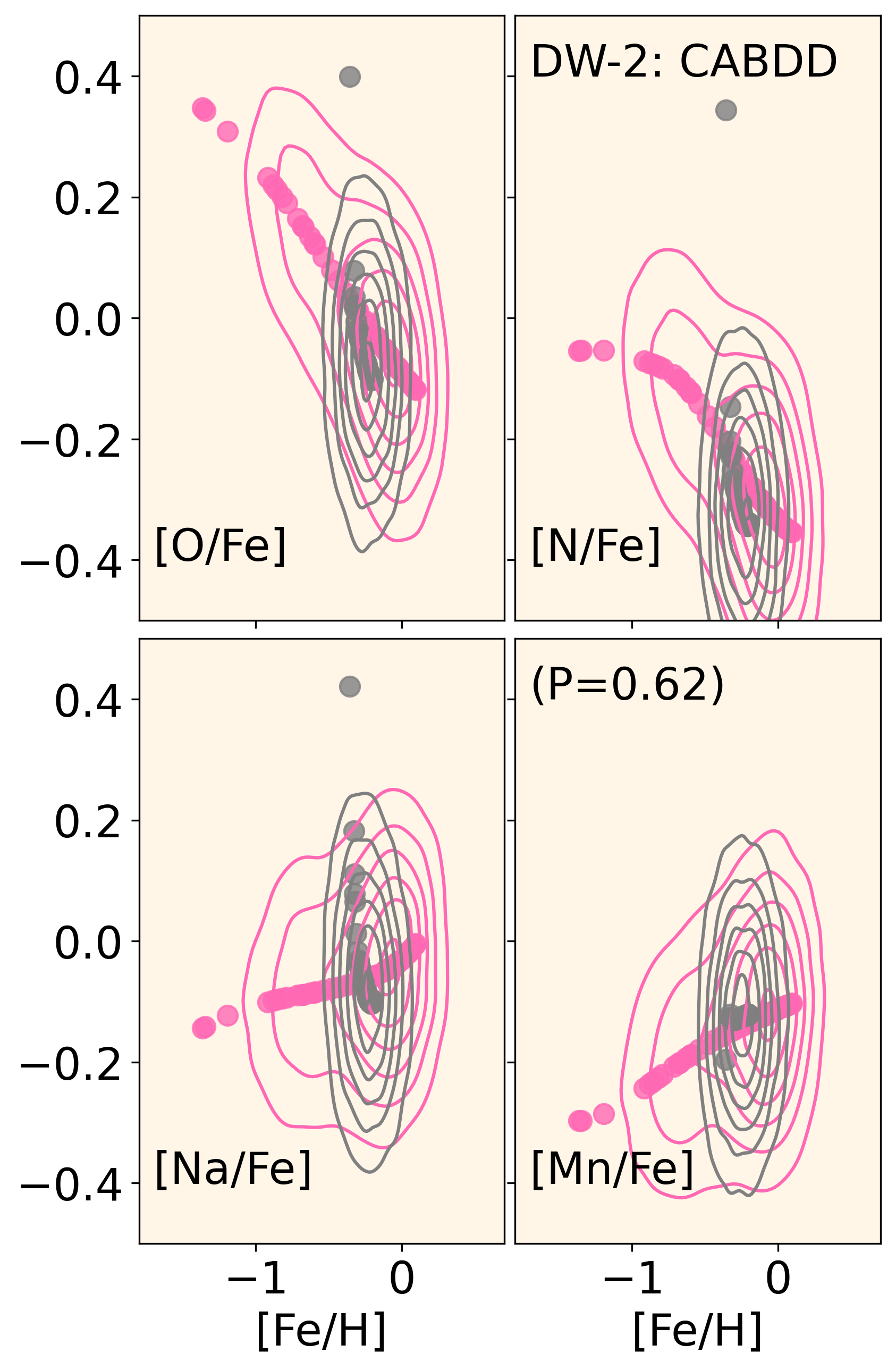}
\includegraphics[width=0.192\linewidth]{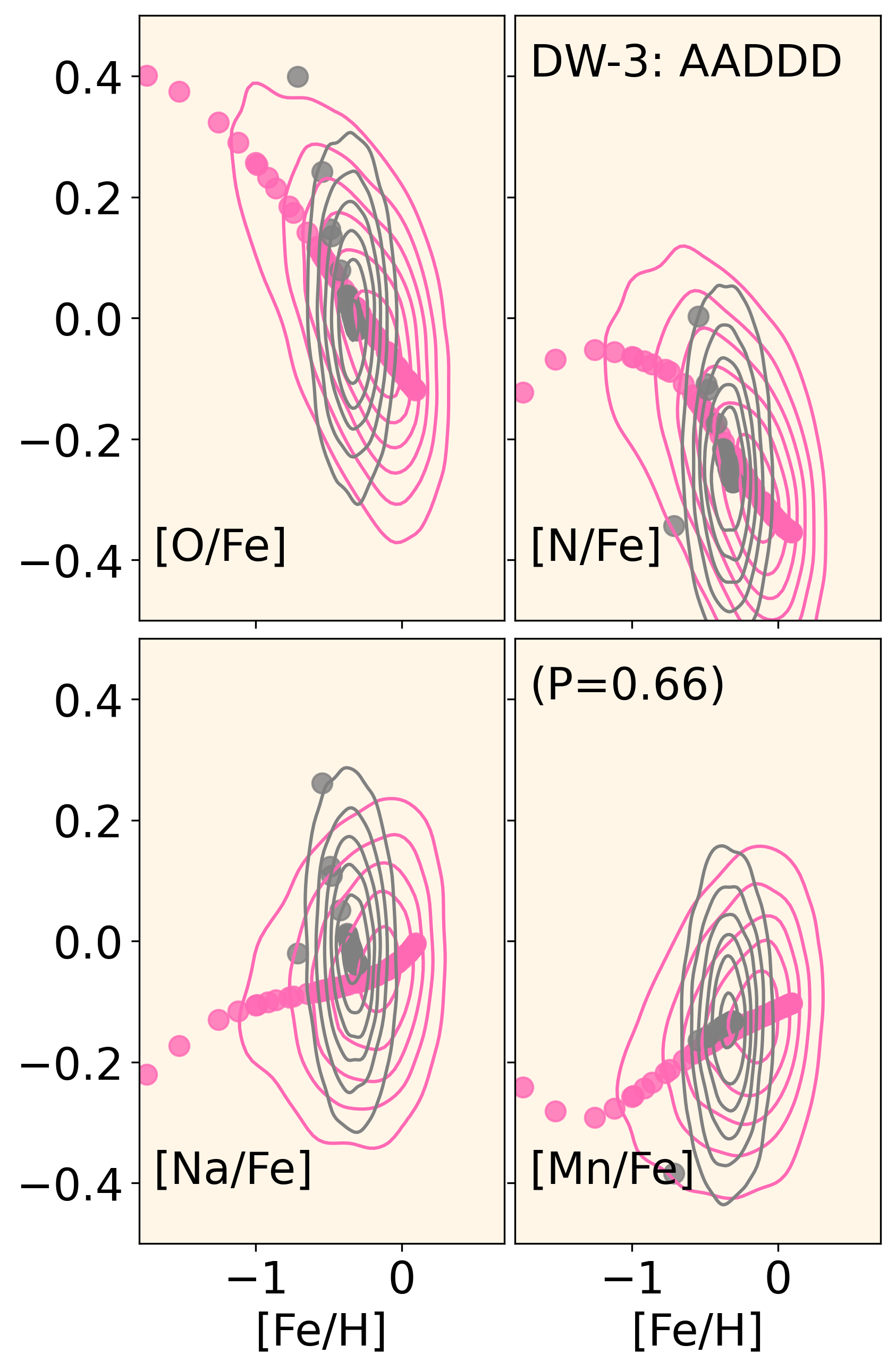}
\includegraphics[width=0.192\linewidth]{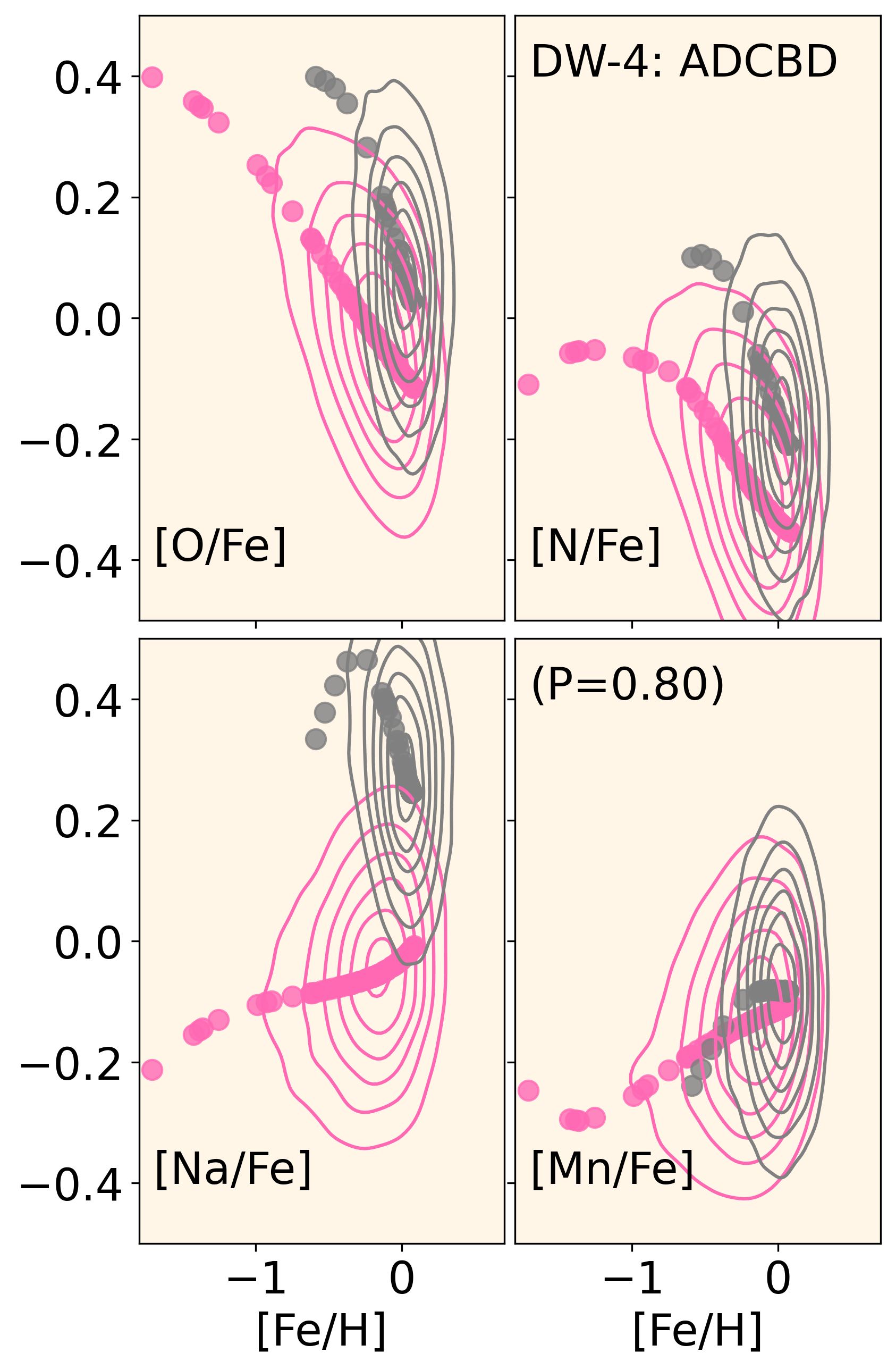}
\includegraphics[width=0.192\linewidth]{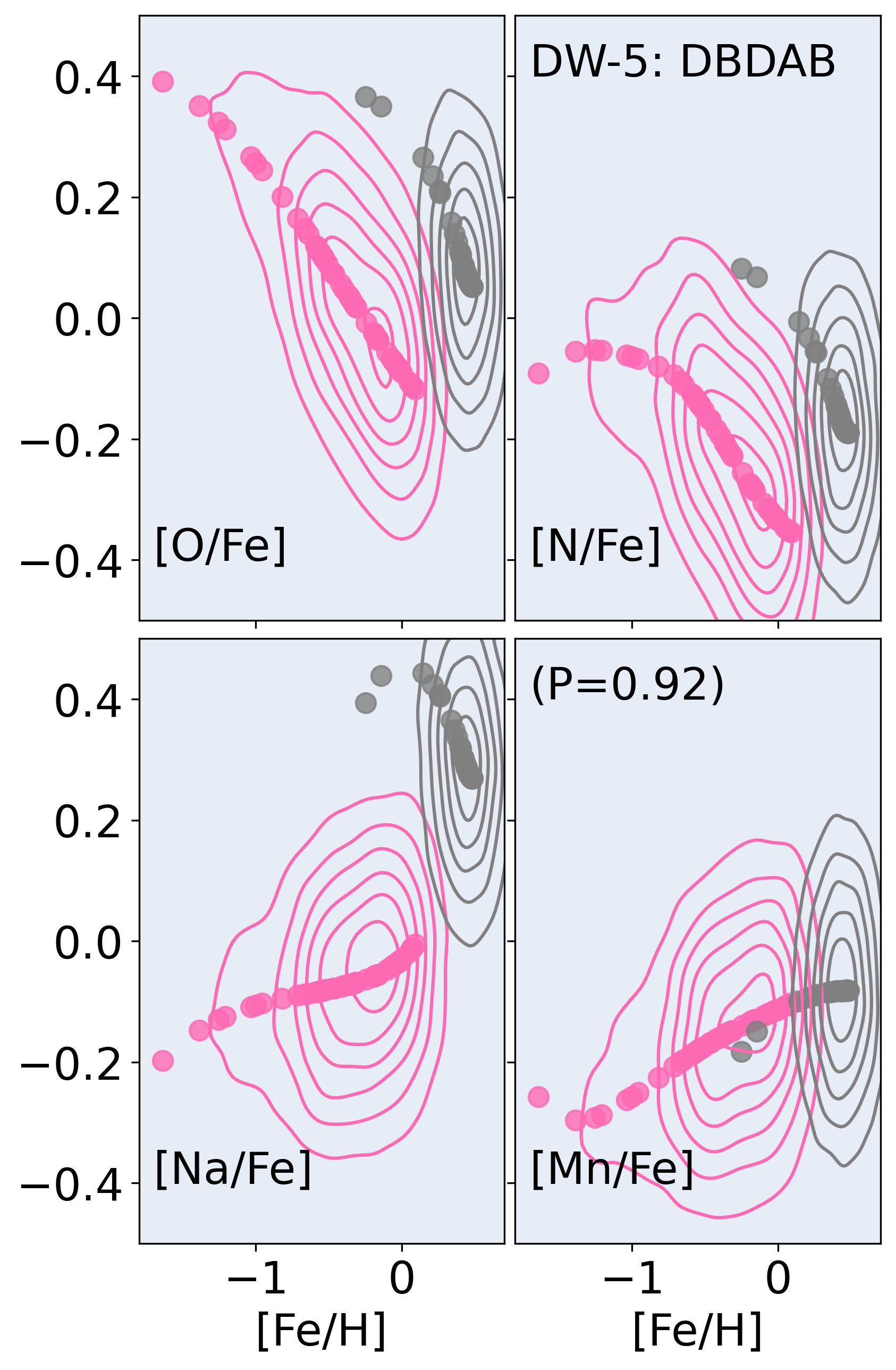}
\caption{Abundance planes for the example models shown in Fig.~\ref{fig:examples_015noise_100}. Green corresponds to the \mwfid\ model, pink to the \dwfid\ model, and gray the grid model. The background color corresponds to the purity of the tree separation. Contours illustrate the trends of the abundance planes considering an error of 0.15~dex. }
\label{fig:planes}
\end{figure*}

\subsection{Case scenarios}\label{sect:cases}

Figure~\ref{fig:examples_015noise_100} shows examples of trees for five different grid models that either do or do not separate from the fiducial models. Purities are indicated. The deepest bipartition is highlighted with a purple circle, enabling where the tree is divided to be seen and thus the purity to be understood. 

The scale of all trees in each diagram is the same, thus allowing direct comparison of the branch lengths between models. Even though we see that trees have two branches, the shape of these branches can vary. Some branches are quite short and tight (e.g., the gray branch of the {\tt ADCBD} model shown in Panel DW-4 of Fig.~\ref{fig:examples_015noise_100}), while other branches are quite long (e.g., the upper branch of the {\tt ABACA} model shown in Panel DW-1). There are models that produce trees with both long and short branches, such as model {\tt DBDAB} shown in Panel 5. Thus, phylogenetic trees might be useful as a clustering method for finding groups and populations, but the topology of the branching pattern could additionally shed light on the evolutionary history of the population. This is discussed further in Sect. \ref{sect:ga}.

Because the trees are made from chemical abundances only, we pay special attention to the four abundance planes previously discussed in Fig.~\ref{fig:alpha_odd_z} for the examples presented in Fig.~\ref{fig:examples_015noise_100}. The trends are plotted in Fig.~\ref{fig:planes}. Background colors follow the purity cut, which is indicated in the title for reference. The colors of the dots and the contours also follow if the trend is from a grid model (gray), or the fiducial models. The top row corresponds to the \mwfid\ model and the bottom row corresponds to the \dwfid\ model.  We draw contours to illustrate the data distribution for uncertainties of 0.15~dex. 

As expected, when the bulk of the abundance distribution of each model differs, the tree separates well. We note that the {\tt ABACA} model has $P\sim0.8$ in both the \mwfid\ and \dwfid\ case, which indicates a poor separation for our classification. Here, despite the distributions being mostly distinct, there is overlap for a few data points. These are the same data points which get shuffled in the tree branches, lowering the purity. While lowering slightly our purity threshold $P$ would reclassify these cases as well separated, we find that this would not affect our overall conclusions over the relative impact of each parameter on branch separation (see Appendix~\ref{app:p08}). 

It is interesting to note that the overlap of the abundance planes drives the confusion in the tree, whereas the difference in overall trend does so to a lesser degree. Also, for tight distributions, a small shift seems to be sufficient for the trees to separate the branches well. For example, for MW-2 and MW-3, the trees have very short branches, but the grid models are still well separated from the \mwfid\ model. 

By relating the abundance planes to the trees, we further realize that the more extended the range in the chemical abundance ratios is, the longer the branches in the tree. This is expected, since the NJ algorithm to build trees uses the difference in chemical abundances to calculate the branch lengths. If differences are large (because the range in abundances is extended), then branches will be large. We recall that our trees are computed considering [X/H] in the abundance ranges. The abundance planes of Fig.~\ref{fig:planes} help us to see the trends of one abundance ratio with respect to another, but not to attribute differences in these planes directly to a distance matrix for the NJ algorithm.

\begin{figure}[t]
    \centering
    \includegraphics[width=1\linewidth]{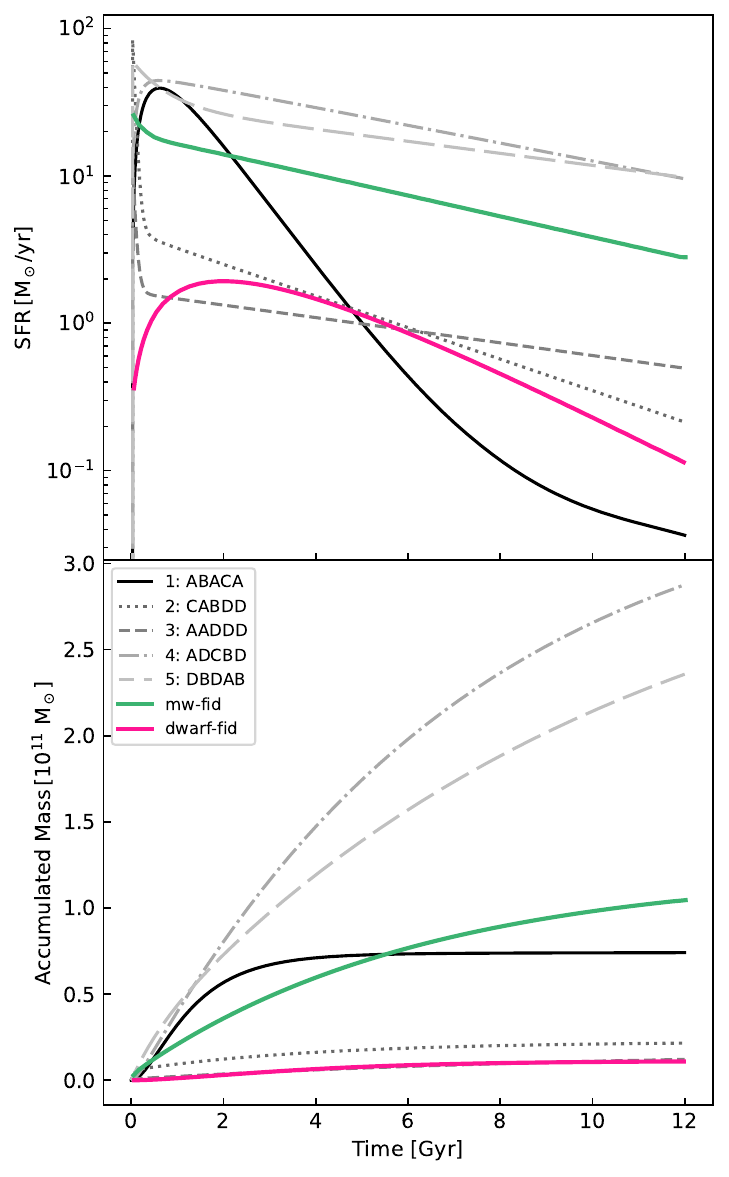}
    \caption{SFR and mass budget as a function of time for the example models shown in Fig.~\ref{fig:examples_015noise_100}.}
    \label{fig:sfr}
\end{figure}

\subsection{Star formation history}

In Fig.~\ref{fig:sfr} we plot the SFR of the five grid model examples displayed in Fig.~\ref{fig:examples_015noise_100}, as well as the fiducial models in the top panel, and the accumulated mass as a function of time in the bottom panel. These plots help us understand the different star formation histories of our models and link them to the different abundance planes and in turn to the trees. 

The first model, {\tt ABACA}, is plotted with a black continuous line. {\tt ABACA} represents a small galaxy because its {\tt A}$ = M_0 = 0.5$ is small. Its inflow of mass is also relatively low ($B = M_1 = 3$) and happens at a high rate ($A = \tau_1 = 1$), the outflow is relatively high ($C = \eta = 3.5$), and its star formation efficiency is low ($D=\nu=0.6$). The SFR of this model is thus high at the start but decreases sharply. The accumulated mass increases fast (due to the low $\tau_1$), but stays more or less constant (due to the combination of high $\eta$ and low $\nu$). The mass budget of {\tt ABACA} has a similar order of magnitude to that of the \mwfid\ model, and its SFR spans a greater range than both fiducial models. This means that, when looking at the abundance planes, there is no overlap with the fiducial models, but there is a continuous distribution. In consequence, there is a slight mixing of the nodes which decreases the purity to $P = 0.8$.

The second example model, {\tt CABDD}, is plotted with a dotted line in Fig.~\ref{fig:sfr}, and represents a galaxy with a high initial mass ($C=M_0=3.5$), even higher than the \mwfid\ model, but with a low inflow mass ($A=M_1=0.9)$, a moderate inflow timescale ($B=\tau_1=4$), a high outflow ($D=\eta=5$), and a high star formation efficiency ($D = \nu = 2.4$).  Its initial SFR is thus very high but decreases sharply during the first Gyr and stays decreasing at high rate during its entire evolution, reaching values that are only slightly above the \dwfid\ SFR. The low inflow but high outflow, combined with the high star formation efficiency, make {\tt CABDD} accumulate mass at a rate similar to that of the \dwfid, despite having a significantly higher initial mass. The evolution of this galaxy translates into abundances that do not change much over time. Because of the significantly different SFR scale, the abundance planes occupy a smaller area than that of the \mwfid. However, the similar mass budget over time with the \dwfid\ makes their abundance planes overlap. As such, even though the SFR is different at the start of the evolution due to different $M_0$, a SFR and mass budget of similar order of magnitude over time do generate abundance planes that overlap and thus mixed phylogenetic branches. This is why {\tt CABDD} does not separate from the \dwfid\ model, but does separate from the \mwfid\ model. 

The third model, {\tt AADDD}, plotted with bold dashed line in Fig.~\ref{fig:sfr}, has a small initial mass budget ($A=M_0=0.5$), small inflow mass ($A=M_1 = 0.9$), slow timescale ($D=\tau_1 = 10$), but very high outflow loading parameter ($D=\eta=5$) and high star formation efficiency ($D = \nu = 2.4$). This model has a similar SFR evolution to the \mwfid\ model but at a scale similar to the \dwfid\ SFR. Furthermore, its accumulated mass is very similar to the \dwfid\ model. This translates into chemical evolution pathways that cross with the \dwfid\ model, causing a low purity for this case. The different scales with the \mwfid\ model make the abundance planes differ, resulting in a phylogenetic tree with two well separated branches.  

Regarding {\tt ADCBD}, plotted with a dashed-dotted line in Fig.~\ref{fig:sfr}, its initial mass is low ($A=M_0=0.5$) but its inflow mass is very large ($D=M_1=9$, $C=\tau_1=7$), and so is its star formation efficiency ($D=\nu=2.4$). The outflow of this model is relatively low ($B=\eta=2$). This translates into a very high SFR over the entire evolution, and a large accumulated mass which increases significantly with time—similar to the \mwfid, but at a higher scale. Thus, the abundance planes of {\tt ADCBD} overlap with the \mwfid\ model but differ from the \dwfid\ model, and so the phylogenetic tree can distinguish with high purity the \dwfid\ model but not the \mwfid\ model. 

The last example, {\tt DBDAB} is a galaxy with a very large initial mass ($D=M_0=5$) but a relatively low mass inflow that arrives at a slow timescale ($B=M_1=3, D=\tau_1=10$). Its outflow is low ($A=\eta=0.5)$, and its star formation efficiency is also moderately low ($B= \boldsymbol\nu=1.2$). This translates into a very high initial SFR that decreases slowly with time, and a mass budget that steadily increases with time. The large difference in mass budget and the different SFR result in abundance planes that are different from both fiducial models and therefore a phylogenetic tree with two branches and high purity. Even though the differences in SFR and mass accumulation are not very large compared to the {\tt ADCBD} case discussed before, we note in Fig.~\ref{fig:examples_015noise_100} that this change in SFR does impact the tree shape, because we see long branches followed by short branches in the tree. This different shape does favor a separation with high purity. 

To summarize, we do not find a direct relation of the scales and shapes of the SFR or the accumulated mass with the tree topologies or the abundances planes. The SFR and accumulated mass can evolve in similar ways because of a combination of input parameters, but abundance planes can differ. The NJ algorithm is able to generate trees with two branches for two models with high purity when abundance planes have a scale that is different in both models. These abundances become different when the rates of change and the scales in the SFR and the accumulated mass are different.  Moreover, if the branches have different shapes due to different SFRs, the NJ algorithm might be able to separate different evolutionary histories even if the abundance planes are not so different. 

The phylogenetic trees are insensitive to the absolute accumulated stellar mass. If $\eta$, $\nu$, and $\tau_1$ remain fixed, changing the overall mass scale modifies only the amplitude of the star formation history, not the chemical pathway followed in abundance space. Since the trees are built exclusively from chemical information, accumulated mass differences are a consequence of the governing parameters rather than the primary driver of branch separation.

\section{Discussion}\label{sect:discussion}

\subsection{The role of outflow in the chemical enrichment history of galaxies}

The outflow mass-loading parameter $\eta$, defined as the ratio between the mass outflow rate and the SFR, quantifies how much interstellar gas is expelled per unit stellar mass formed. Among all the free parameters explored, $\eta$ is the one that most consistently drives a separation between phylogenetic branches (Fig.~\ref{fig:shap}). 
One reason for this is that $\eta$ directly regulates how difficult it is for a galaxy to retain and accumulate gas. A larger $\eta$ implies that a greater fraction of the ISM is removed each time stars form \citep{Marlowe95, Rhee26}. As a consequence, the gas reservoir is more strongly regulated and less capable of sustained accumulation. This affects not only the total gas content, but also how effectively newly produced metals remain in the system \citep{Gibson13, Xu22}.

The effect $\eta$ has on the chemical evolution of the models is not immediate \citep{andrews17}, but it alters the relative location of the [$\alpha$/Fe]–[Fe/H] knee \citep{Weinberg17}. A higher value of $\eta$ means that less of the $\alpha$ elements produced by massive stars early on in the evolution will be retained, directly altering the chemical makeup of a given model in its later stages. Indeed, we can see in Panels 2 and 3 of Fig.~\ref{fig:planes}, which plot grid models with high $\eta$ in gray (final letter in the model name is {\tt D}), that the [O/Fe] values remain low at all metallicities and that the metallicity never increases as much. Because phylogenetic trees are constructed purely from chemical abundances, the impact $\eta$ has on the chemical trajectory of a galaxy translates into a structural effect on the final tree.  It alters the pathway followed in abundance space rather than merely rescaling the evolutionary timescale.

In contrast, we have the star formation efficiency $\nu$, which controls how rapidly gas is converted to stars by the relation described in Eq. \ref{eq:sfr}. As such, while it can affect the speed at which chemical evolution occurs, it does not shape the overall trajectory of chemical evolution \citep{Vincenzo17}. This causes $\nu$ to have a less direct impact on the final structure of the tree.  Even though the trees are built using [X/H], the different timescales of enrichment of the different elements considered make that each [X/H] evolves at different rates. Thus, the position of the $\alpha-$knee both in [Fe/H] and in [$\alpha$/Fe] has an impact on the tree topology and thus the purity when combining models with different $\alpha-$knees.

We can see this play out when combining with a dwarf-like system, such as \dwfid, where $\eta$ is already large and star formation is inefficient. This allows $\nu$ to become the second most important parameter controlling branch separation.
Low $\nu$ implies that only a small amount of stellar mass forms at early times. Consequently, the build-up of iron from CCSNe is slow, and the metallicity reached before Type Ia supernovae become important remains low. This shifts the knee in [$\alpha$/Fe]–[Fe/H] toward lower metallicities (see, e.g., the [O/Fe]-[Fe/H] planes of Fig.~\ref{fig:planes}, Panels 1 and 4). In this regime, enrichment is limited by inefficient star formation rather than by prolonged gas dilution, which explains why $\nu$ is particularly relevant for dwarf comparisons.

When combining models to Milky Way–like systems (such as the \mwfid\ one), the second most influential parameter is the inflow timescale, $\tau_1$. This parameter determines how long primordial gas continues to enter the galaxy and dilute the interstellar medium. A longer inflow timescale prolongs the dilution phase, slowing the rise in metallicity and shifting the knee toward lower [Fe/H]. In such systems, enrichment is not primarily limited by inefficient star formation but by sustained dilution from gas accretion. Therefore, $\tau_1$ becomes more important than $\nu$ in differentiating chemical trajectories.

\begin{figure*}[t]
\centering
\includegraphics[scale=0.4]{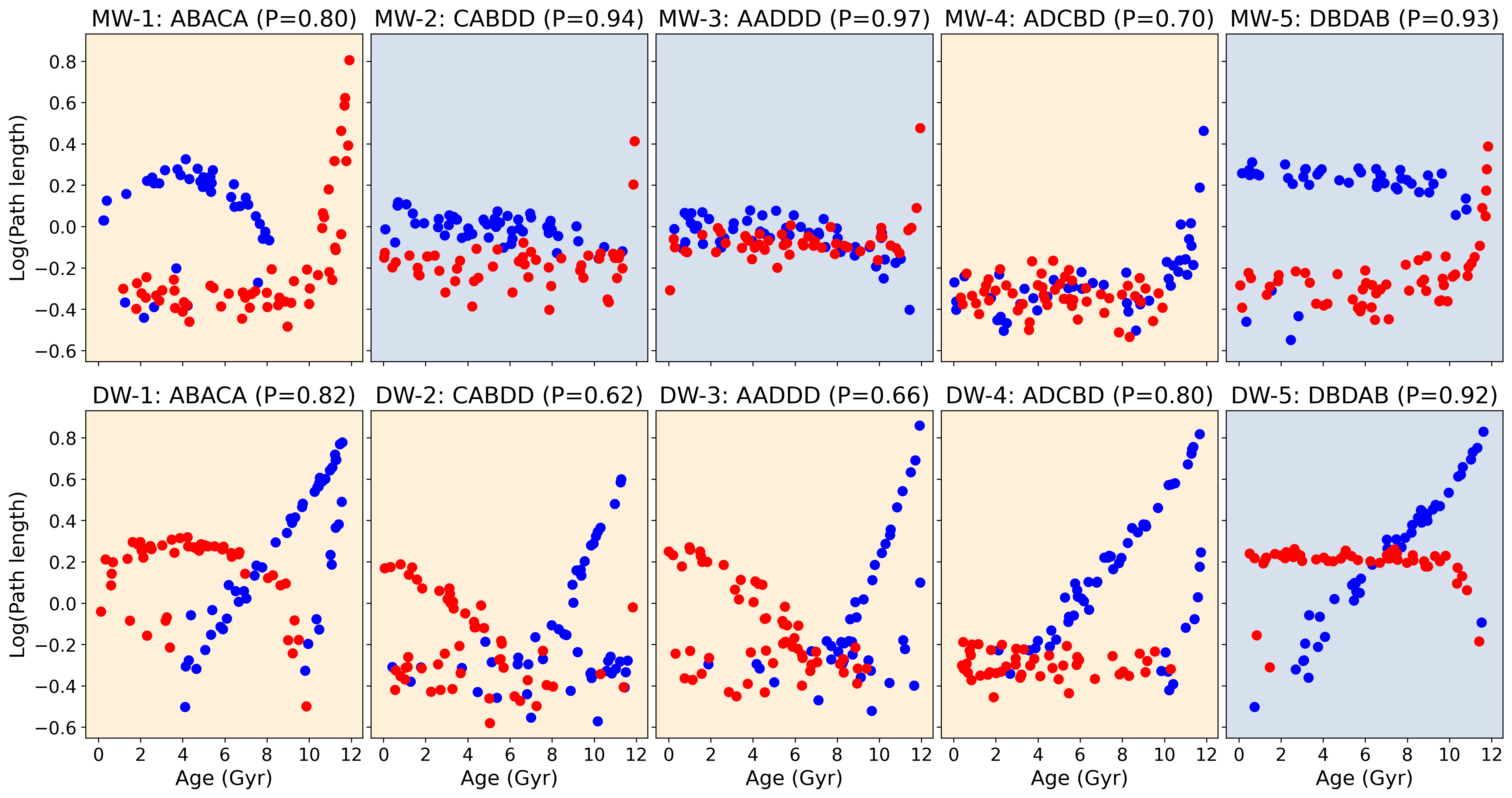}
       \caption{Path length in log scale from tip to central bifurcation as a function of age for the trees shown in Fig.~\ref{fig:examples_015noise_100}, with the background colors following that same figure. Each branch is represented with a different color. }  
   \label{fig:lengths} 
\end{figure*}

\subsection{Applications to Galactic archaeology}\label{sect:ga}

Galactic archaeology is the field of using stellar atmospheres, in particular the chemical composition imprinted in the spectra, as fossil records for galaxy evolution. Today, millions of spectra from which stellar parameters and chemical compositions can be determined with high precision have become available \citep{GaiaEso, Galah, Apogee}. Combined with Gaia, we know the kinematics of all these stars \citep{GaiaDR3}. The information about fossil stars is thus extremely rich and diverse and is only increasing \citep{2019Msngr.175....3D, 2026AJ....171...52K}. 

However, in order to reconstruct the history of the Milky Way, we do not only need high-precision data of stars in large quantities, we also need efficient methods to extract the information from this data. If the hypothesis of Galactic archaeology is that the chemical composition of low-mass stars has information of the evolution of the Milky Way, then we expect that we can chemically tag stars and find the building blocks of the Galaxy \citep{Freeman2002}.  This has motivated the development of clustering algorithms for chemical tagging. 

Strong chemical tagging aims at finding clumps in chemical space only to reconstruct the open clusters in the disk. But this task faces challenges. The degeneracies in chemical space, which translate into stars of similar chemistry but record a different history (see also the overlap in chemical space of some models in Fig.~\ref{fig:planes}), in addition to uncertainties in the measurements, still hamper an accurate and complete clustering of the chemical data \citep{2025A&A...702A.267S, 2024A&A...688A.165S, 2021A&A...654A.151C}. 
Weak chemical tagging intends to generalize this problem by adding further information to the chemistry, such as ages and kinematics, which enables some of the degeneracies to be broken down \citep{2025A&A...704A..40T, 2025ApJ...985..129B, 2025A&A...704A.296Y} and galactic components to be identified. But ages are not always known accurately, and kinematical memory can be erased with time, so applications of weak chemical tagging remain limited. 

Moreover, in both of these cases there is a conceptual limitation, as clustering stars by similar chemistry cannot select stellar populations with prompt chemical evolution. Indeed, stellar populations should not clump in chemistry or age. When clumps are found by clustering algorithms, putting them together in a historical context remains extremely challenging as the result of the clustering does not immediately tell us how the found clusters are related to each other \citep{2018A&A...619A.125A, Meszaros21}. 

Phylogenetics offers a promising complementary approach to chemical tagging. First, visualizing the data in trees instead of clumps enables us to find stellar families that have evolved as branches that allow for change in chemistry. In this study, from the abundance planes of Fig.~\ref{fig:planes}, a classical clustering is not always as able to distinguish the evolutionary pathways as the trees, because it groups the data of both models with similar chemical compositions together (see Appendix~\ref{app:gmm}). Standard clustering algorithms in particular will fail to assign stars with an extended metallicity distribution to a single group. While the NJ algorithm does not always separate the evolutionary pathways, it does so when the chemical evolution of the models, in addition to producing different chemical trends, lead to branches that have different topologies. 

In Fig.~\ref{fig:lengths} we show the path lengths of the trees displayed in Fig.~\ref{fig:examples_015noise_100}. We plot each tip as a function of age in each branch by a different color, following \citep{Jofre2017}. The length is calculated to the bifurcation shown by the purple circle in Fig.~\ref{fig:examples_015noise_100}. We can see that the branches can have a variety of topologies as the tips evolve. Some have sharp changes at old ages, others remain approximately constant, others increase steadily, and yet others increase at young ages. In general, we see that the oldest tips have the largest paths, which is consistent with the fact that the branches of the models join at the most metal-rich stars, since this is when the models become most similar to each other.  We can finally see that trees with low $P$ (orange background) do not show clear trends of path length with age. In these cases, one can see that the histories between the models are confused in a single branch (see, e.g., Panel DW-1 of Fig.~\ref{fig:lengths}).  

Lastly, trees give us the missing information of how the different groups might be related to each other, and how the history of these groups is shared. It is interesting that the most metal-rich, or youngest, stars are those that connect the two main branches. If all of these models had started with pristine gas, then the early stage of evolution in the models could have been much more abrupt than that of later times. Even in this wide range of chemical evolution conditions, the chemical abundances evolve to similar values with time, decreasing the diversity of the chemical data. This is already observed in the Milky Way, where the very ancient disk stars have a higher dispersion in metallicity compared to the young and metal-rich disk \citep{2008A&A...480...91S, 2014A&A...562A..71B}. Regarding elemental abundances, \citet{2016A&A...593A.125S} shows that the old stars in the disk have a larger spread in abundances compared to the young stars. Indeed, chemical abundance gradients in the disk traced by open clusters also have shown that the gradients are flatter towards younger ages, i.e., as time passes, stars formed in different parts of the disk tend to be chemically more homogeneous  \citep{2023A&A...669A.119M}.

However, this is remains in a way opposite to biological trees, which increase in diversity as they evolve off their common ancestor, and thus the phylogenetic trees have been designed as an intuitive tool to study this increase of diversity with time. Because descent with modification is present in galaxy evolution, investigating this difference offers exciting opportunities for future research regarding evolutionary behavior.

\section{Summary and conclusions}\label{sect:conclusions}

This paper presents the first phylogenetic study using analytical models of chemical evolution. Inspired by the fact that galaxies are a collection of stars representing a fossil record of a myriad of histories, our goal was to investigate how these methods can help disentangle chemical evolution histories under different scenarios. To this aim, we generated a set of analytical models with \flexce\ \citep{andrews17} that allowed us to compute the evolution of chemical abundances for different initial masses ($M_0$), mass inflows ($M_1$), mass inflow timescales ($\tau_1$), mass outflow loading factors ($\eta$), and star formation efficiencies ($\nu$). 

Different chemical enrichment paths result when different input parameters are adopted. However, not all differences may be significant if we consider typical uncertainties in the abundance measurements of observations, which might be on the order of 0.15 dex for some elements \citep{2019ARA&A..57..571J, 2025AJ....170...96M, 2025PASA...42...51B}. Thus, in this paper we sought to identify the input parameter ($M_0, M_1, \tau_1, \eta, \nu$) that drives a sufficient difference in the chemical pattern such that we can identify different branches in a phylogenetic tree that records a specific history 

We therefore built phylogenetic trees combining a random selection of 100 tips from two sets of models, a fiducial one and one from a grid from Table~\ref{tab:grid}. Using the classical neighbor-joining algorithm \citep{Saitou1987}, which is based on the distance matrix calculated from the total difference of all [X/H] abundances in the models, we created phylogenetic trees. To automatically assess if these trees had separated branches that reflect a given chemical evolution model, we defined a purity parameter ($P$) that corresponds to the averaged percentage of tips of a specific model in each branch.

Because of the large parameter range to explore and the interdependency of such parameters, we determined which parameter was the most important by running random forests. This process was used to classify the purity considering a threshold of $P>0.9$ for a well-separated branch and $P\leq0.9$ for a poorly separated branch. 
 
Our results showed that $\eta$, namely the outflow mass-loading parameter, has the highest impact when using phylogenetic trees to distinguish different evolutionary histories. This result was independent of the adopted fiducial model.  
Specifically, trees separated best when $\eta$ had a very different value between the combined models. This result is consistent with the literature, which has shown that outflows have a large impact on the chemical compositions of galaxies \citep{Gibson13}. Indeed, we could see how different $\eta$ parameters would produce different scales and positions of the $\alpha-$[Fe/H] knee, which in turn would produce different branch topologies and enable a better separation with the NJ algorithm. 
 
The role of $\eta$, or any other input parameter, can be seen in the SFR or the mass accumulation of the models (Fig.~\ref{fig:sfr}). However, the shape and scale of the SFR or total mass do not directly predict the trends of the abundance planes. It is the combination of all parameters, but the role of $\eta$ most importantly, that impacts the overall chemical evolution of the models. Thus, if we aim to apply the NJ algorithm to disentangle stellar population observations of Milky Way stars, we might have more success if these populations differed significantly in their outflows. 

Applying our technique to Milky Way stars or to cosmological simulations would require further investigation of how to extend this work towards combinations of more than two models. Yet, even in this simple first step, we were able to recognize the power of phylogenetic techniques. While clustering algorithms are limited to finding groups with small diversities, phylogenetic techniques allow us to disentangle groups with extended distributions. Our comparison of Gaussian mixture models versus NJ algorithm presented good evidence. 
Phylogenetics therefore represents a powerful tool because with trees, not only can we find stellar families, but we can also study how their histories might be related. 

The fact that the different models in this work connected in the NJ tree at high metallicities deserves attention. While this is expected from a perspective of galaxy evolution, where metal-poor stars are more distinct from each other than metal-rich stars \citep{2023A&A...669A.119M}, this is in a way contrary to the expectations from biological systems, where diversity increases with time. Our result offers new opportunities to understand the paths of evolution in different environments. Stars inherit metals from their progenitors, just as humans inherit DNA from their ancestors, and so descent with modification occurs in galaxies.

\section*{Data availability}
The fiducial and grid models generated and analyzed in this work can be downloaded from Zenodo (\url{https://zenodo.org/records/22833949}). The data analysis routines to were adopted from the Python libraries {\tt sklearn} and {\tt BioPython}.

\begin{acknowledgements}
      We thank the anonymous referee for their helpful comments, which significantly improved the methodology presented in this work. R.C. and P.J. acknowledge support from ANID InES G\'enero INGE230004. This is a project of the Phylogal (\url{www.phylogal.cl}) collaboration. We thank Álvaro Rojas-Arriagada, Matías Neto, Pablo Villarreal, and Francisco Cubillos for illuminating discussions throughout the development of this work. We acknowledge financial support from FONDECYT Regular Grant Number 1231057. C.A.G. acknowledges support from ANID through FONDECYT Regular 1262342.
\end{acknowledgements}

\bibliographystyle{aa}
\bibliography{references}

\appendix

\section{Abundance planes of the fiducial models}\label{app:planes}

\begin{figure}[h] 
   \centering
   \includegraphics[width=3.3 in]{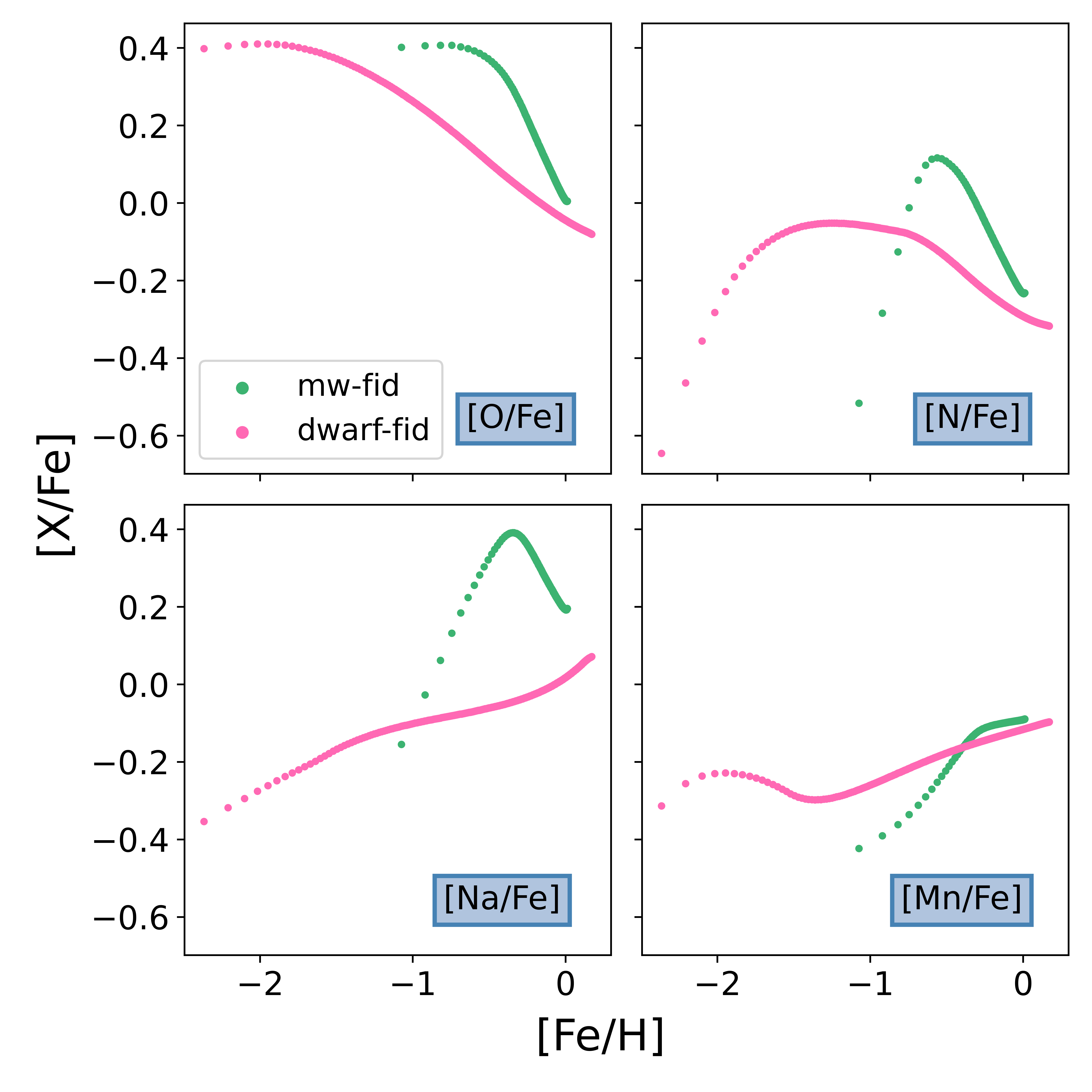}
   \caption{Abundance ratios as a function of metallicity for the \mwfid\ model (green) and the \dwfid\ model (pink). The abundances from Fig. \ref{fig:abund_time} are shown here.}
   \label{fig:alpha_odd_z}
\end{figure}

Figure~\ref{fig:alpha_odd_z} shows the chemical distribution of the same elements discussed in Fig.~\ref{fig:abund_time}, but this time the abundance ratios are with respect to Fe and plotted against [Fe/H]. We plot in each panel one of the elements discussed above, with green representing the \mwfid\ model and pink the \dwfid\ model. 
It is possible to see that the relationships between different elements change as [Fe/H] increases, which is due to the different production rates. For example, while [Na/Fe] has a steady increase with [Fe/H] for the \dwfid\ model, for the \mwfid\ model the ratio has a steep increase until $\mathrm{[Fe/H]} = -0.3$, at which point the ratio starts decreasing. 

It is important to note the relative abundance of Na in these models, given that it dominates over the other abundances for a significant fraction of the evolution. This is shared across the \flexce\ models considered in this work, though it does not necessarily reflect the chemical abundance pattern of a Milky Way-type galaxy.

\section{Phylogenetics with Python}\label{app:trees}
Because in astronomy no phylogenetic model for galaxy evolution is available, we use the classical ``distance-based" method \citep{lemey09} to build phylogenetic trees. This method considers that the larger the distance between leaves in a tree, the more distant these leaves are in their evolutionary history. 

The very first step is to define a distance metric for calculating the pairwise distance between all of the objects in the system. In our case, the objects are data points which represent the chemical composition of the ISM at a certain time of the {\tt flexCE} model. We consider all chemical element ratios [X/H], calculate their Euclidean distance and save it in a distance matrix, which is symmetrical and its diagonal is zero. More specifically, we use the Python routine called {\tt DistanceMatrix} from the {\tt BioPython} library. 

The second step is to build the tree, which in this case is done with the neighbor-joining algorithm \citep{Saitou1987} calling the Python routine {\tt DistanceTreeConstructor.nj}, also from {\tt BioPython}. The output is a tree array that, with the help of various routines from {\tt BioPython}, allows us to save, plot, or navigate the tree. The latter is for extracting information about the tips, the nodes, the splits, the branch lengths, etc. 

In this work, we perturbed the distance matrix by adding noise to the data and calculated several trees for the same model by also randomly selecting 100 data points, which allowed us to have an estimate of the uncertainty of the purity parameter. 

The standard format for saving trees is the {\tt newick} format, which is a string array. This format allows us to open and further analyze trees with other widely used computer packages and programming languages in biology.

\section{Random forest results}\label{app:rf}

\subsection{Input and output of the random forest model}
We called the Python {\tt RandomForestClassifier} routine of {\tt sklearn}. Our classification is binary, 0 for when $P<0.9$ and 1 for when $P>0.9$.  The model considers five input features, which are $M_0$, $M_1$, $\tau_1$, $\eta$ and $\nu$. These features  have the values listed in Table~\ref{tab:grid}. The random forest model uses 500 decision trees with a depth of five nodes. The confusion matrices in this case are  
\[
\begin{bmatrix}
170 & 1 \\
24 & 113
\end{bmatrix}
\] for the \mwfid\ model, and 

\[
\begin{bmatrix}
 244  &  1\\
  5 & 58
\end{bmatrix}
\] for the \dwfid\ model. In this case, the \mwfid\ model had an accuracy of 94.34\% $\pm$ 1.89\%, and the \dwfid\ model had an accuracy of 97.75\% $\pm$ 1.46\%. Other hyperparameters for all RF models considered in this work can be found at the end of this appendix.  Finally, to find the Shapley values, we use the Python routine {\tt shap.TreeExplainer} inputting the random forest.

\subsection{Considering a cut of $P=0.8$}\label{app:p08}
Here we present the input parameter analysis for our tree sample, considering a purity threshold of $P=0.8$, rather than the value of $P=0.9$ adopted previously. For the same noise of 0.15 dex, we find that 763 \mwfid\ and 576 \dwfid\ trees separate under this new criteria. We redid the random forest 
analysis, following the same procedure described in \ref{sect:mapping} and using the same routines as explained above. The confusion matrices and accuracies in this case are 
\[
\begin{bmatrix}
64 & 15 \\
4 & 225
\end{bmatrix}
\] and 96.39\% $\pm$ 1.8\% for the \mwfid\ model, and 

\[
\begin{bmatrix}
 149  &  7\\
  3 & 149
\end{bmatrix}
\] and 94.04\% $\pm$ 3.2\% for the \dwfid\ model. Other hyperparameters can be found in Table~\ref{tab:hyperparams}.

\begin{figure*}[h] 
\centering
\includegraphics[scale=0.45]{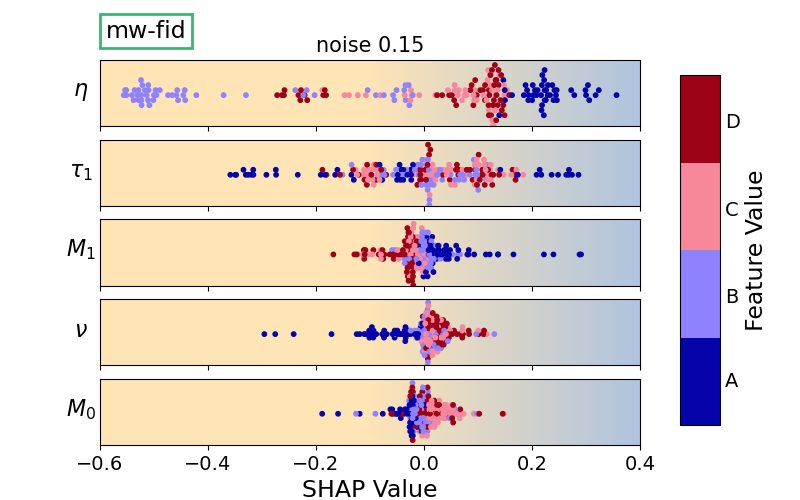}
\includegraphics[scale=0.45]{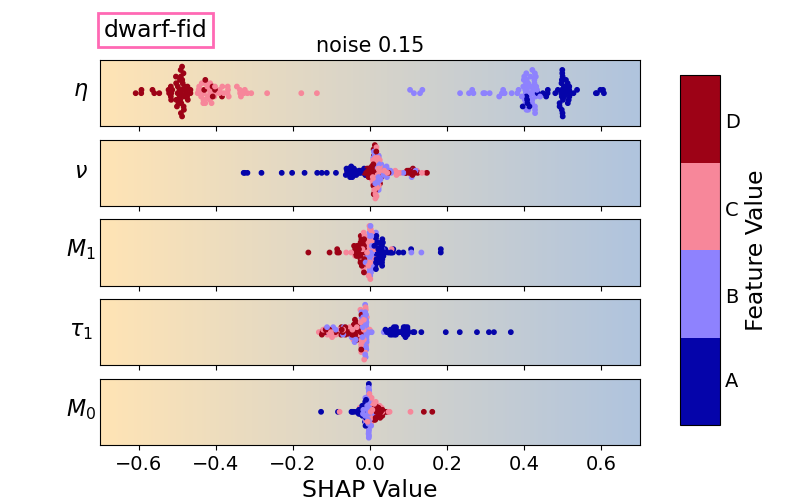}
       \caption{Random forest interpreted with Shapley values for the models compared to the \mwfid\ (\textit{left}) and for the models compared to the \dwfid\ (\textit{right}). A purity threshold of $P>0.8$ and a noise of 0.15 dex are considered.}
   \label{fig:shap08}
\end{figure*}

Starting with the random forests in Fig. \ref{fig:shap08}, we find that for both cases $\eta$ is the most significant parameter when it comes to determining whether a tree separates or not. Additionally, we find that the values farthest from those of the fiducial model contribute the most. This matches the previous results for $P=0.9$. However, we do observe that the Shapley values reach a lower limit for this new threshold, implying that these $\eta$ values have a deeper impact on a given tree not separating.

For the \mwfid\ case specifically, $\tau_1$ remains the second most influential parameter for separation, where the lowest values considered contribute to a tree not separating. It is interesting to note, however, that those same low values are also linked to a higher impact on separation, meaning that this parameter is not as influential for determining tree separation as it was when $P=0.9$.

Regarding the \dwfid\ case, we find that $\eta$ has become far more influential for separation than when $P=0.9$, as seen by the high magnitude of the Shapley values and the relative absence of intermediate values. $\tau_1$ displays a similar behavior for the positive values, while the rest of the parameters have very little bearing on the separation of a given tree. This clustering of the Shapley values goes hand in hand with the more relaxed purity threshold considered in this appendix, since a given value of $\eta$ or $\tau_1$ has a greater opportunity of separating the tree.

\subsection{Considering 200 tips} \label{app:rf_200}
\begin{figure*}[h] 
\centering
\includegraphics[scale=0.45]{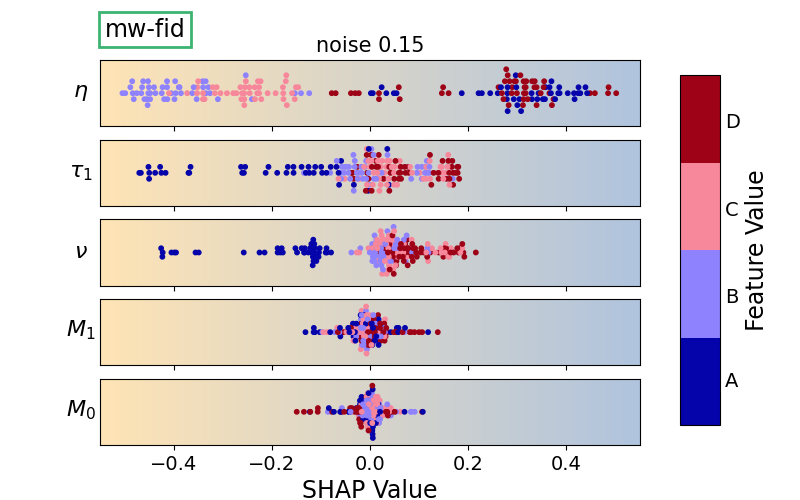}
\includegraphics[scale=0.45]{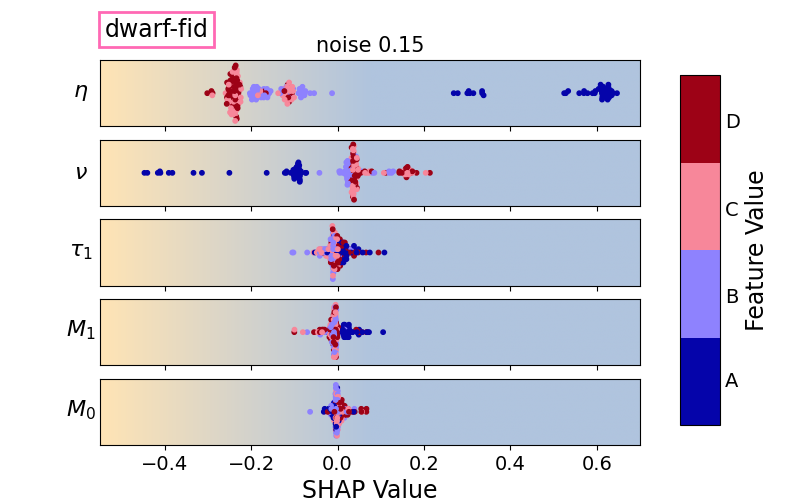}
       \caption{Random forest Shapley values for the models compared to the \mwfid\ (\textit{left}) and for the models compared to the \dwfid\ (\textit{right}). A purity threshold of $P>0.9$ and a noise of 0.15 dex are considered. The trees used here contain 200 tips, as opposed to the 400 used in the main analysis of this paper.}
   \label{fig:shap200}
\end{figure*}

Here we repeat the analysis but select randomly 200 tips instead of 100. Results remain unaffected, as seen in Fig.~\ref{fig:shap200}. The confusion matrices and accuracies in this case are 
\[
\begin{bmatrix}
161 & 4 \\
37 & 106
\end{bmatrix}
\] and 90.72\% $\pm$ 2.56\% for the \mwfid\ model, and 

\[
\begin{bmatrix}
 241  &  1\\
  11 & 55
\end{bmatrix}
\] and 95.9\% $\pm$ 1.67\% for the \dwfid\ model. Other hyperparameters can be found in Table~\ref{tab:hyperparams}. 

\begin{figure*}[h] 
\centering
\includegraphics[scale=0.45]{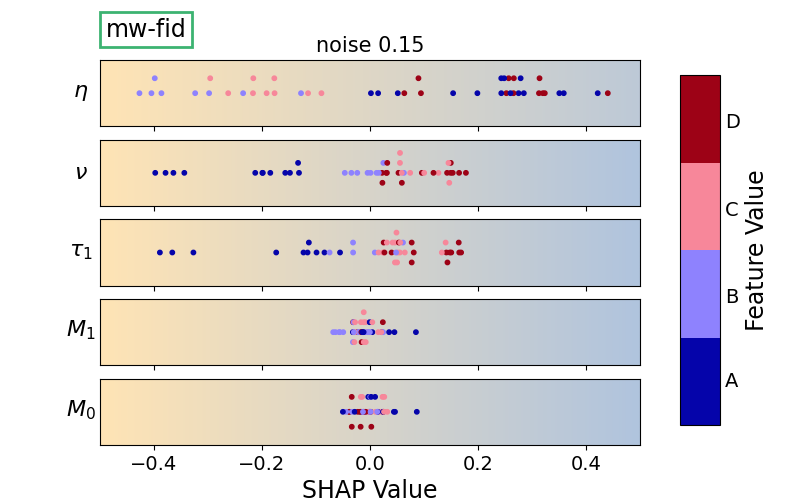}
\includegraphics[scale=0.45]{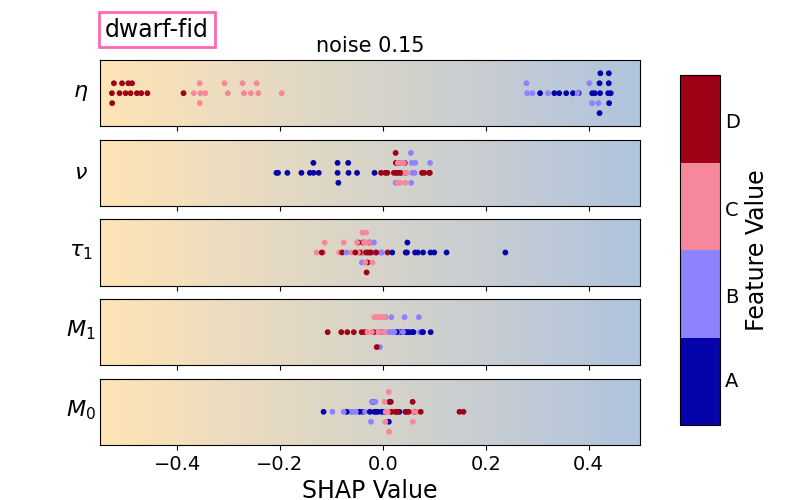}
       \caption{Random forest Shapley values for the models compared to the \mwfid\ (\textit{left}) and for the models compared to the \dwfid\ (\textit{right}). A purity threshold of $P>0.9$ and a noise of 0.15 dex are considered. The trees used here contain 800 tips, as opposed to the 400 used in the main analysis of this paper. Note that each plot only considers a quarter of the total sample (i.e., 256 trees instead of 1024) due to the computational expense that these larger trees entail.}
   \label{fig:shap800}
\end{figure*}

\subsection{Removing Na from the distance matrix}\label{sec:noNashap}
We test our results without [Na/H] in the distance matrix. To do so, we build random trees using all abundances except Na, with random uncertainties of 0.15 dex, and determine the purities. We then study the random forest following the same procedure as above. The confusion matrices and accuracies in this case are  
\[
\begin{bmatrix}
174 & 6 \\
19 & 109
\end{bmatrix}
\] and 97.46\% $\pm$ 1.09\% for the \mwfid\ model, and 

\[
\begin{bmatrix}
 243  &  4\\
  6 & 55
\end{bmatrix}
\] and 97.95\% $\pm$ 1.34\% for the \dwfid\ model. Other hyperparameters can be found in Table~\ref{tab:hyperparams}.

\begin{figure*}[h] 
\centering
\includegraphics[scale=0.45]{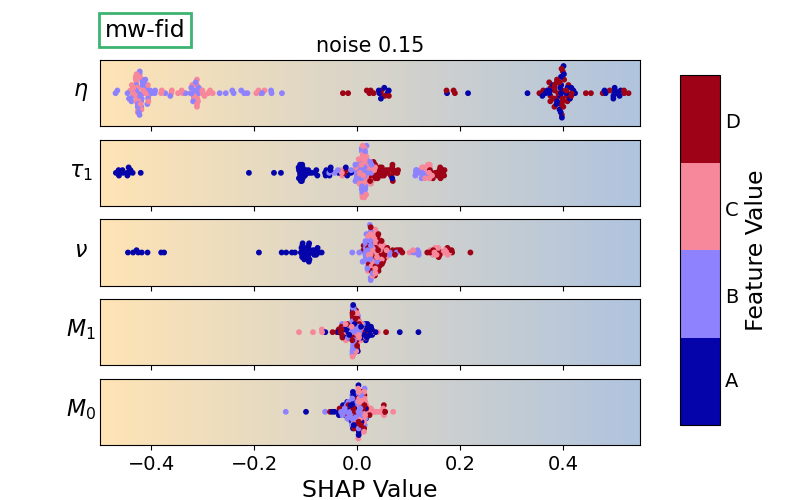}
\includegraphics[scale=0.45]{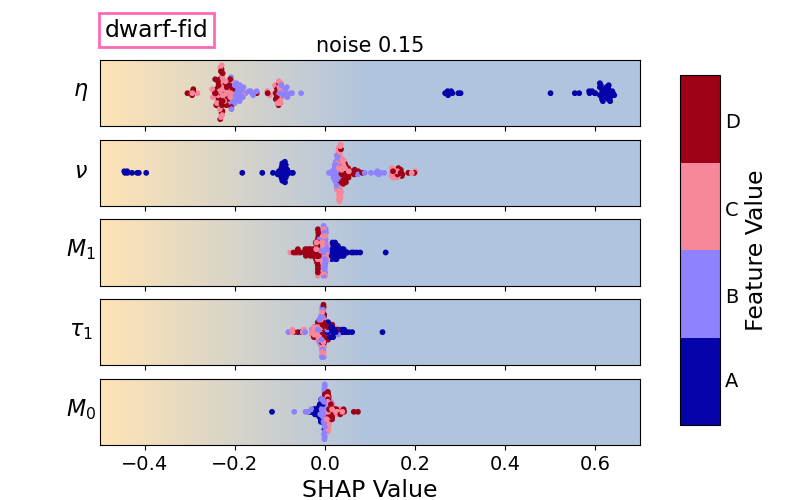}
       \caption{Random forest interpreted with Shapley values for the models compared to the \mwfid\ (\textit{left}) and for the models compared to the \dwfid\ (\textit{right}). A purity threshold of $P>0.9$ and a noise of 0.15 dex are considered. All abundance ratios except [Na/H] have been used for the distance matrix.} 
   \label{fig:noNashap}
\end{figure*}

The results for the \dwfid\ and the \mwfid\ can be found in Fig.~\ref{fig:noNashap}. Without Na, $\eta$ is still the parameter that affects the separation of trees most; however, minor differences are seen in the Shapley values. A larger impact is seen for the \mwfid\ model, which was shown to have very high Na abundances. 

\subsection{Results with uncertainties of 0.08 dex}\label{app:error008}

Here we present the random forest  analyses for our tree sample, considering a noise of 0.08 dex, as opposed to the 0.15 dex adopted previously. Using $P=0.9$ as our purity threshold, we find that 652 \mwfid\ and 326 \dwfid\ separate with these criteria.

Regarding the random forests for both the \mwfid\ and \dwfid\ cases, the overall behavior resembles that of the noisier scenario. The main difference is the increased significance of $\nu$, replacing $\tau_1$ as the second most significant parameter for tree separation. $\nu$ in these two cases becomes more significant when it comes to determining when a tree does not separate, as evidenced by the low Shapley values associated with the lowest feature value. For the other feature values, the impact on tree separation is similar to or smaller than that of other parameters. The relevance of $\nu$ may have been obscured by the higher noise value discussed in the body of the paper, as a higher noise would lead to a decrease in the magnitude of the Shapley values, and as such a drop in the parameter order within the random forest.

The confusion matrices and accuracies in this case are  
\[
\begin{bmatrix}
99 & 13 \\
7 & 189
\end{bmatrix}
\] and 94.73\% $\pm$ 1.75\% for the \mwfid\ model, and 

\[
\begin{bmatrix}
 200 & 10\\
   9 & 89
\end{bmatrix}
\] and 97.36\% $\pm$ 1.38\% for the \dwfid\ model. Other hyperparameters can be found in Table~\ref{tab:hyperparams}.

\begin{figure*}[h] 
\centering
\includegraphics[scale=0.45]{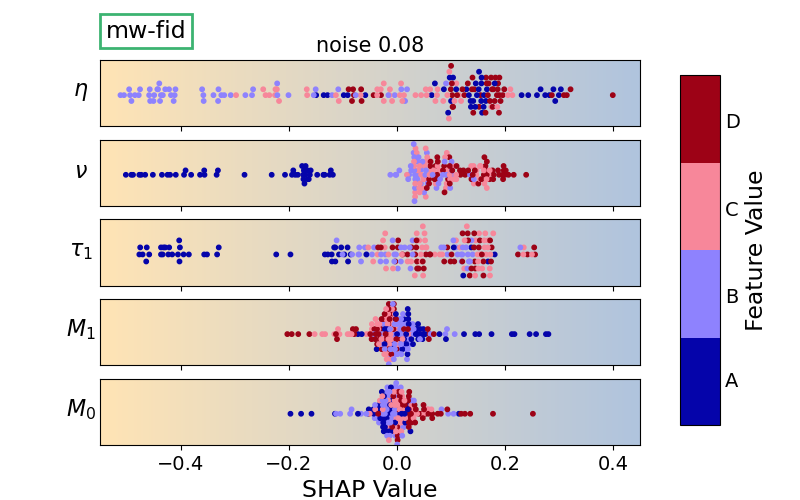}
\includegraphics[scale=0.45]{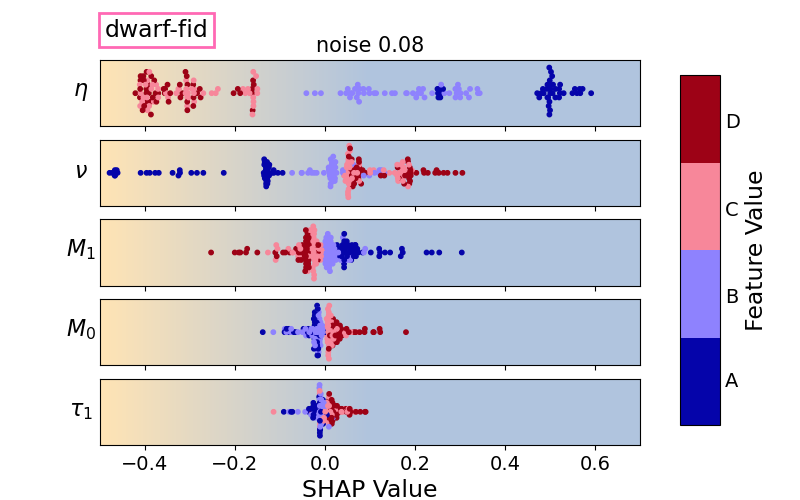}
       \caption{Random forest modified by Shapley values for the models compared to the \mwfid\ (\textit{left}) and for the models compared to the \dwfid\ (\textit{right}). A purity threshold of $P>0.9$ and a noise of 0.08 dex are considered.}
   \label{fig:noise08}
\end{figure*}

\subsection{RF model hyperparameters}\label{app:r_hyper}
In Table~\ref{tab:hyperparams} we list the hyperparameters resulting from the various random forest models used in this work. 

\begin{table*}[h]
    \centering
    \caption{Hyperparameters for various RF models discussed in this work.}
    \begin{tabular}{c | cccc | cccc}
            & \multicolumn{3}{c}{\mwfid} &&  \multicolumn{3}{c}{\dwfid} \\
            \hline 
         class & precision  & recall & f1-score&    support & precision  & recall & f1-score&    support\\
         \hline \\
         
          \multicolumn{9}{c}{ $P = 0.9$ cut and $\sigma=0.15$ dex}\\
         \hline
            0    &   0.88  &    0.99  &    0.93  &    171
                 &   0.98  &    0.99  &    0.99  &    245 \\
           1     &  0.99   &   0.82   &   0.90  &    137
                 & 0.97  &  0.94   &   0.95   &     63 \\

        \hline \\      
          \multicolumn{9}{c}{ $P = 0.8$ cut and $\sigma=0.15$ dex}\\
         \hline
           0   &    0.83   &   0.94   &   0.88   &     79
              & 0.98   &   0.96   &   0.97   &    156 \\
           1   &    0.98  &    0.93   &   0.96  &     229 
           &   0.96   &   0.98  &    0.97    &   152 \\ 

        \hline \\      
          \multicolumn{9}{c}{$P = 0.9$ cut, $\sigma=0.15$ dex, and 200 tips}\\
         \hline

          0     &  0.81   &   0.98   &   0.89   &    165
              & 0.96   &   1.00  &    0.98   &    242 \\
           1    &   0.96   &   0.74   &   0.84  &     143
           & 0.98   &   0.83  &    0.90    &    66 \\    
        \hline \\      
          \multicolumn{9}{c}{$P = 0.9$ cut, $\sigma=0.15$ dex, and without [Na/H],}\\
          
         \hline

          0     &  0.90  &    0.97    &  0.93  &     180
              & 0.98   &   0.98   &   0.98  &     247 \\
           1    &  0.95  &    0.85  &    0.90    & 128
           & 0.93  &    0.90   &   0.92   &     61 \\

                \hline \\      
          \multicolumn{9}{c}{$P = 0.9$ cut, $\sigma=0.08$ dex}\\
          
         \hline

          0     &  0.93    &  0.88  &    0.91  &     112
              & 0.96   &   0.95 &     0.95  &     210 \\
           1    &  0.94    &  0.96  &    0.95  &     196 
           & 0.90    &  0.91  &    0.90   &     98   \\

           \hline

    \end{tabular}
    \label{tab:hyperparams}
\end{table*}

\clearpage 
\onecolumn

\section{Clustering with Gaussian mixture models}\label{app:gmm}

\begin{figure*}[h]
    \centering
\includegraphics[width=0.192\linewidth]{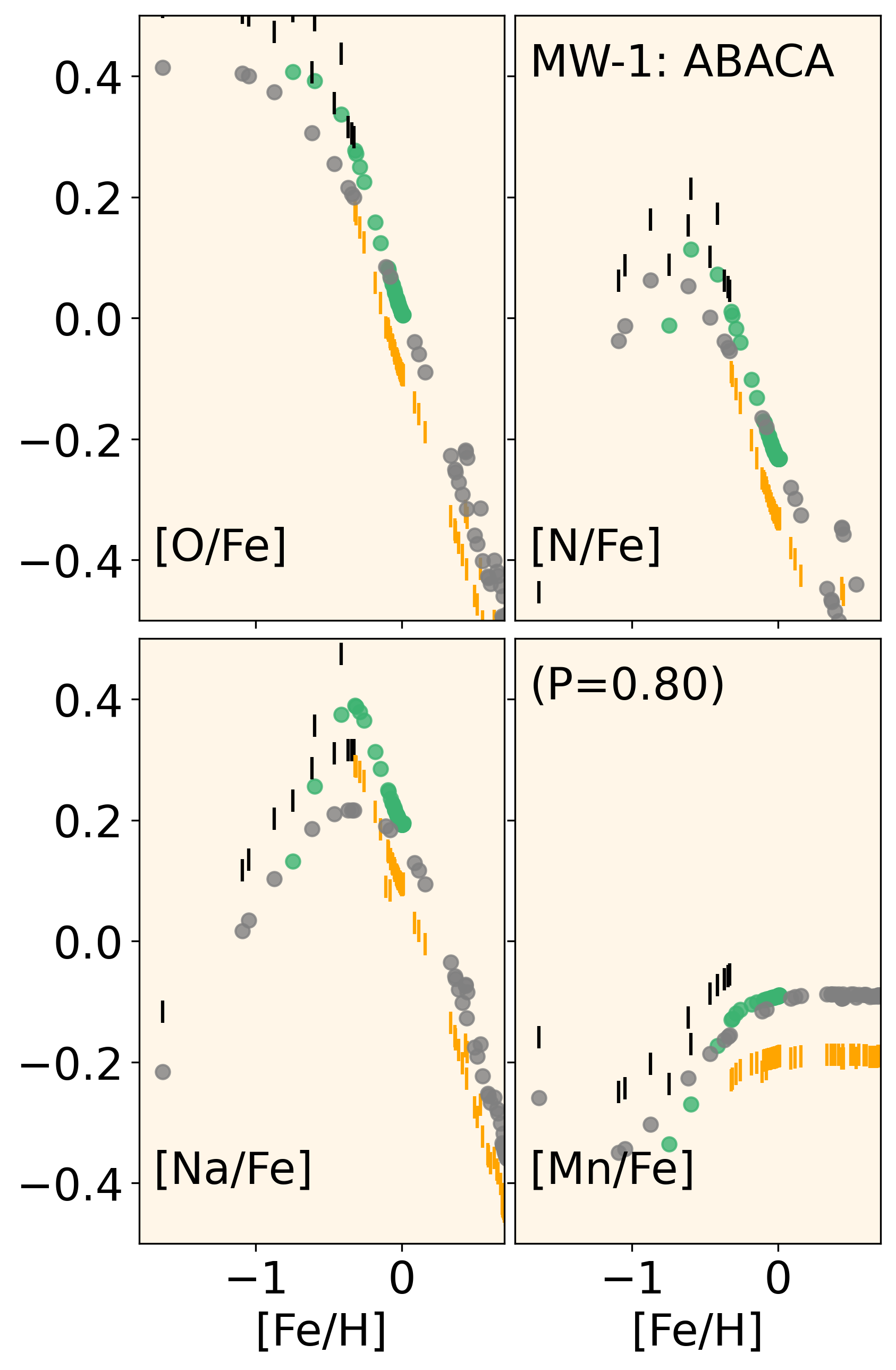}
\includegraphics[width=0.192\linewidth]{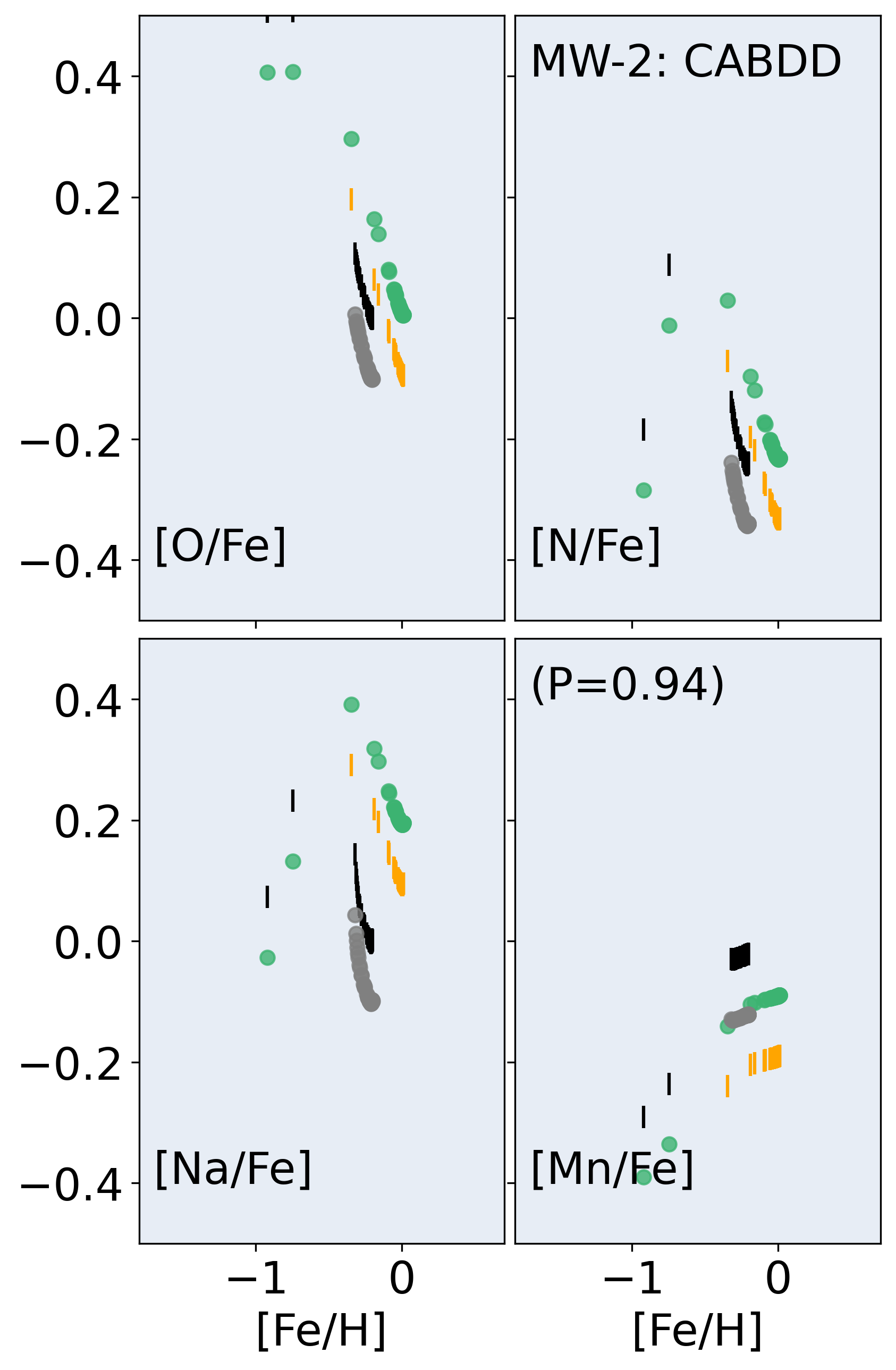}
\includegraphics[width=0.192\linewidth]{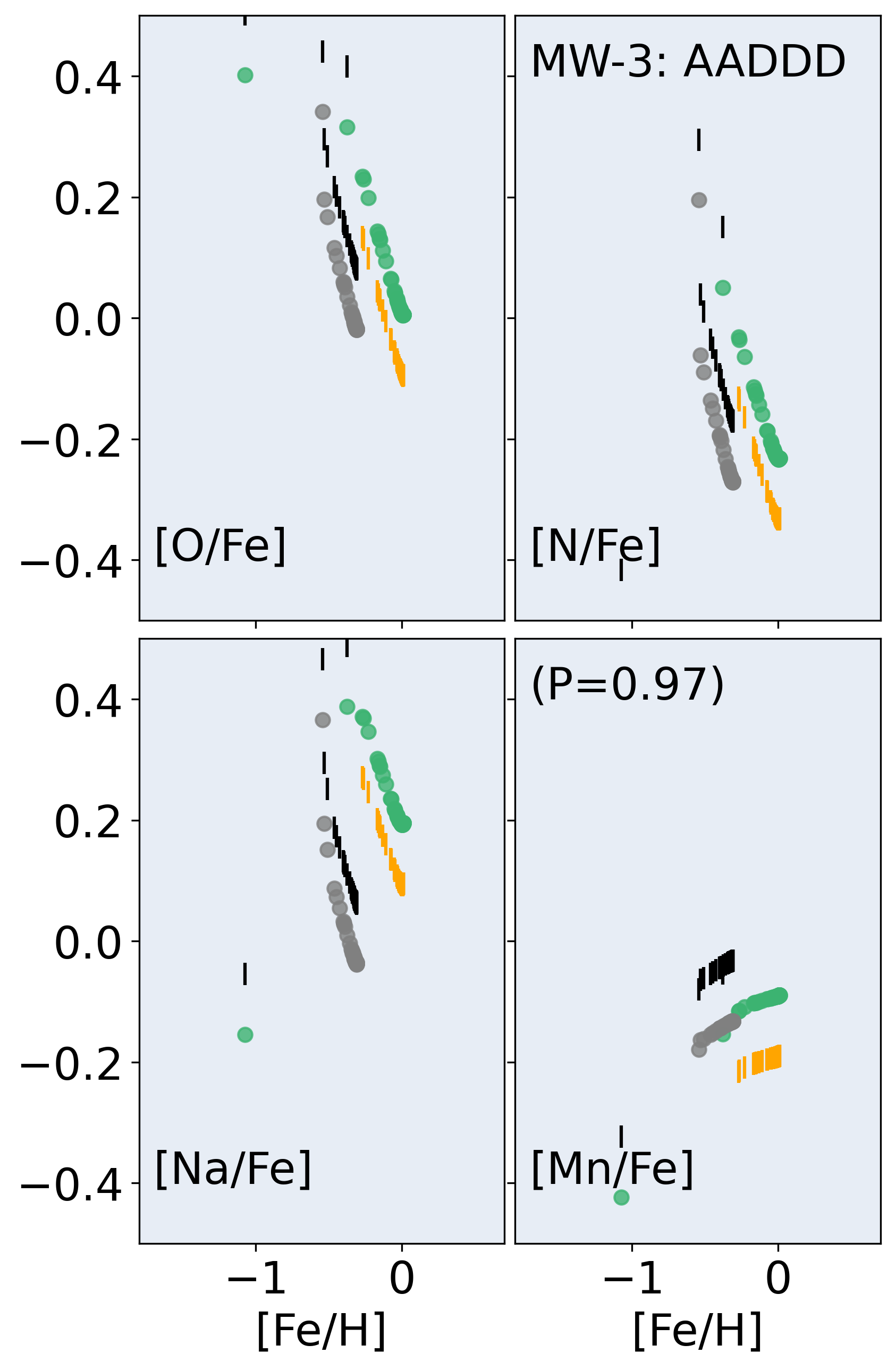}
\includegraphics[width=0.192\linewidth]{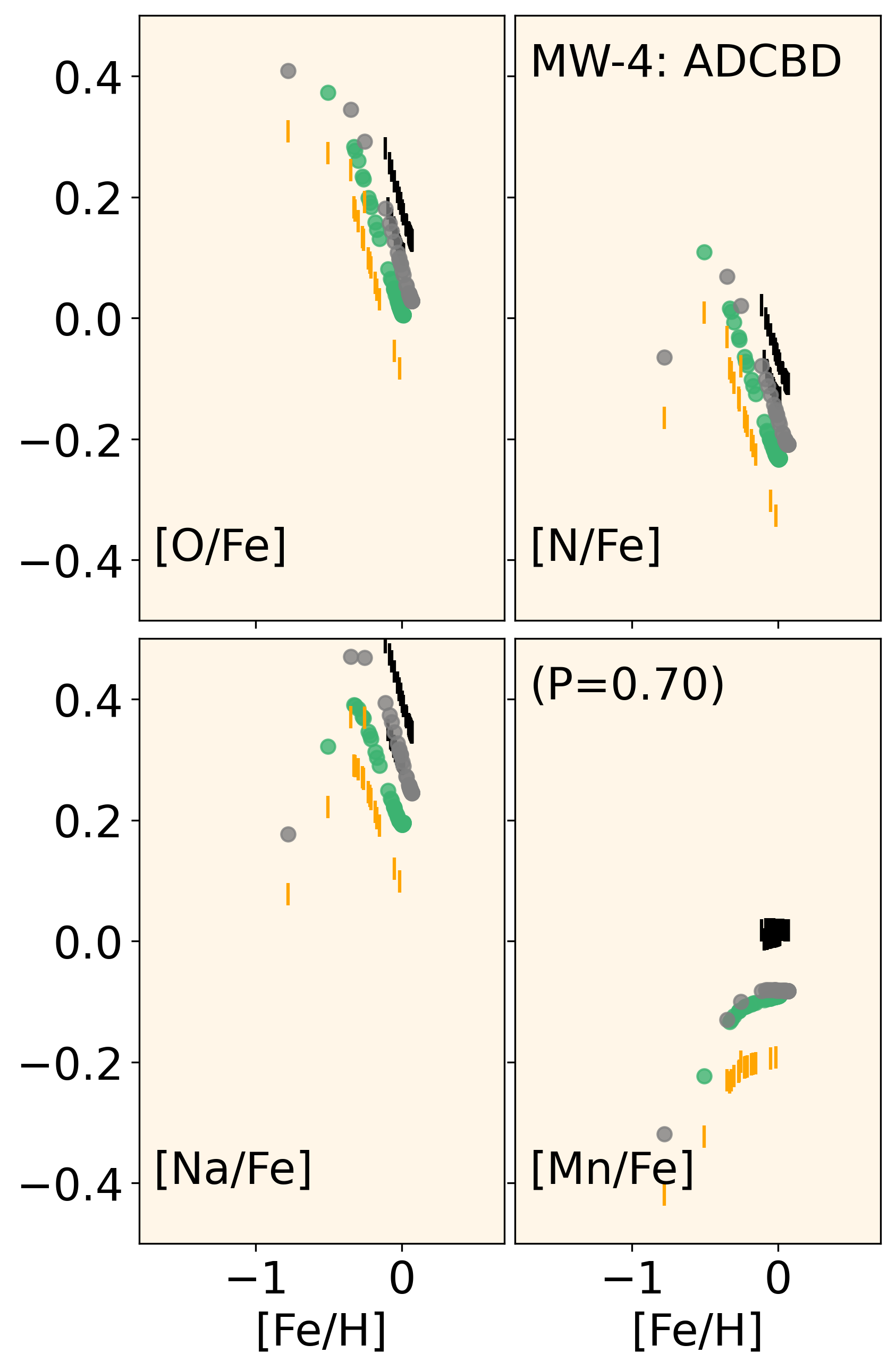}
\includegraphics[width=0.192\linewidth]{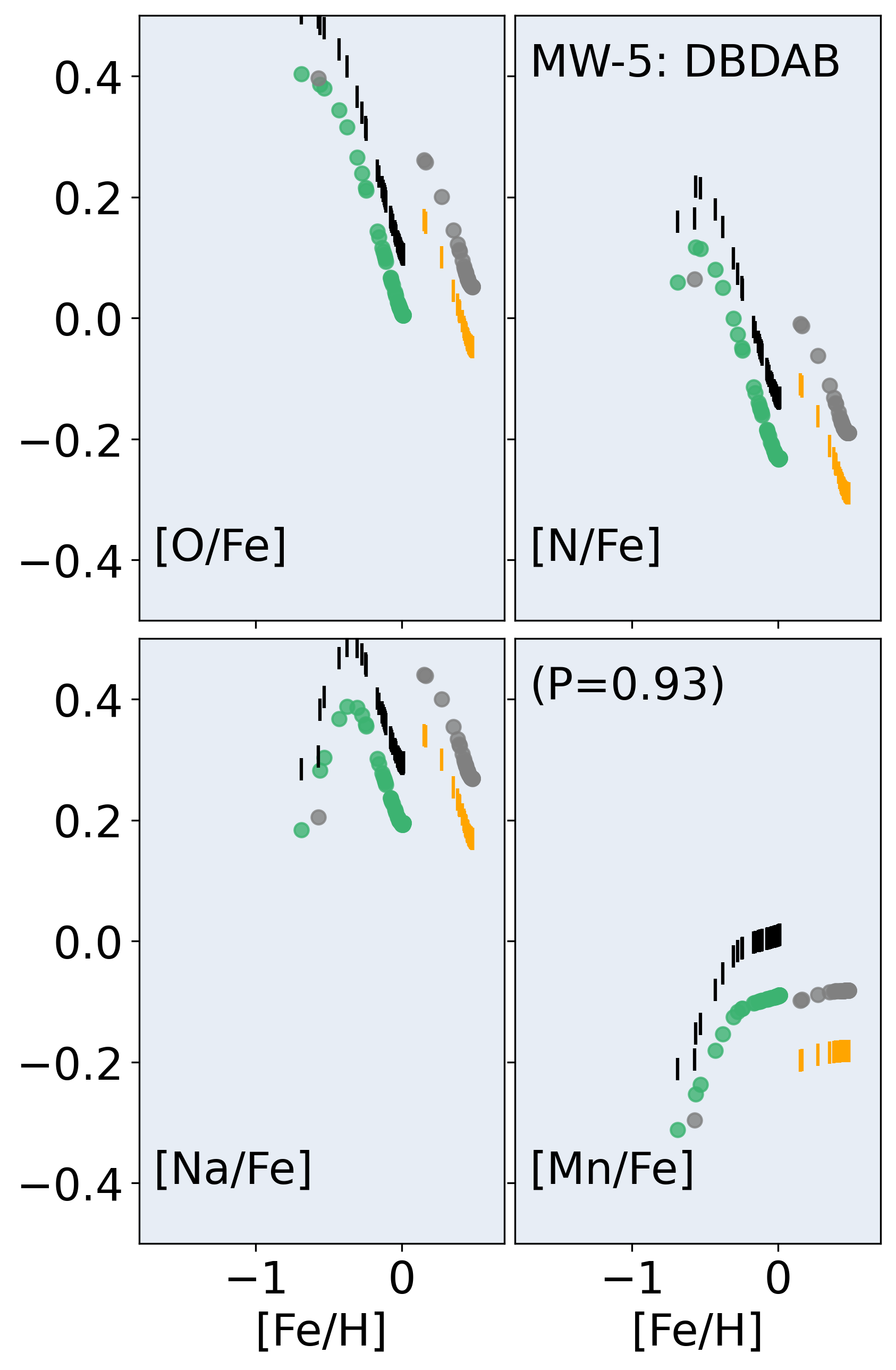}

\includegraphics[width=0.192\linewidth]{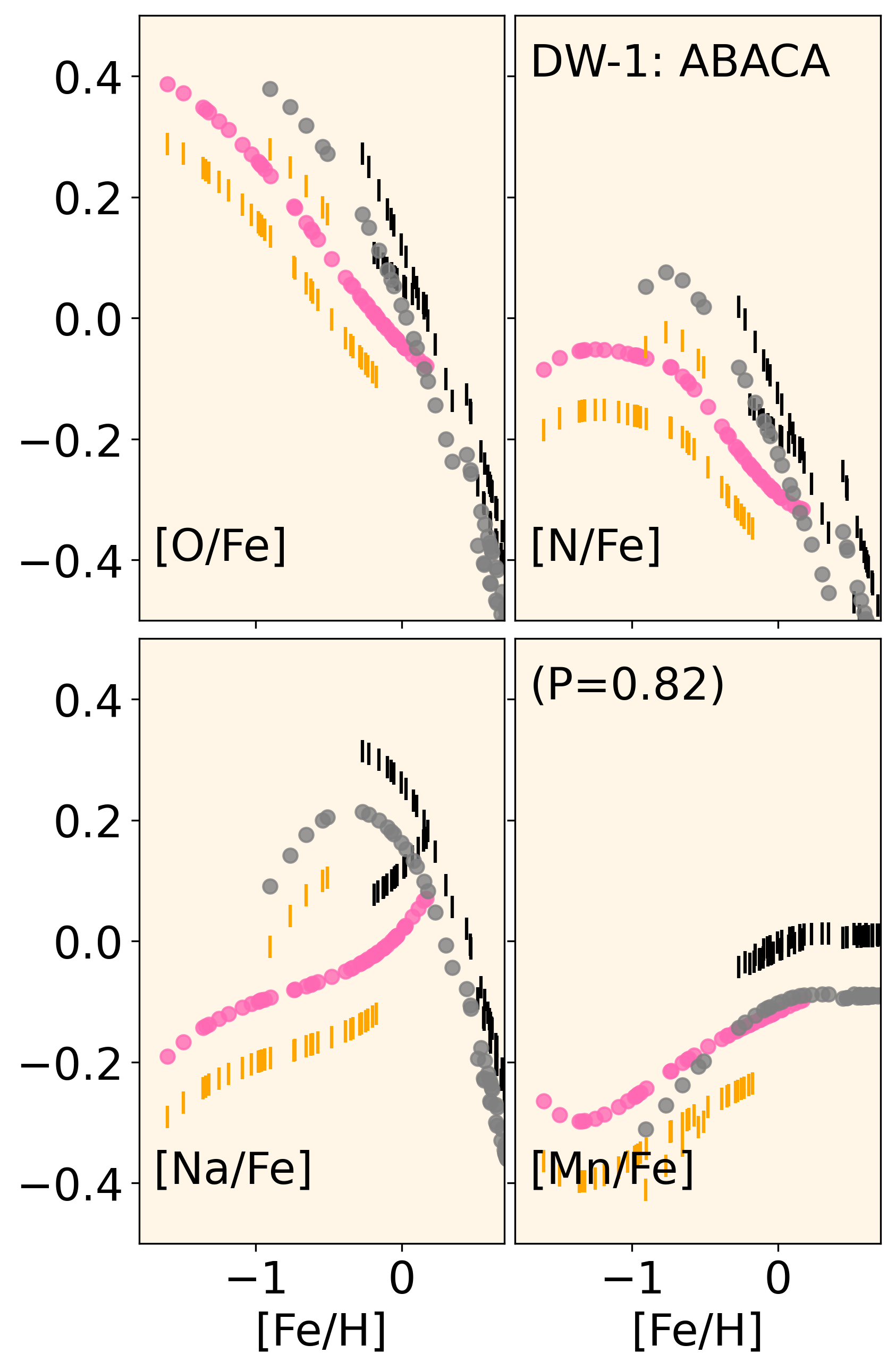}
\includegraphics[width=0.192\linewidth]{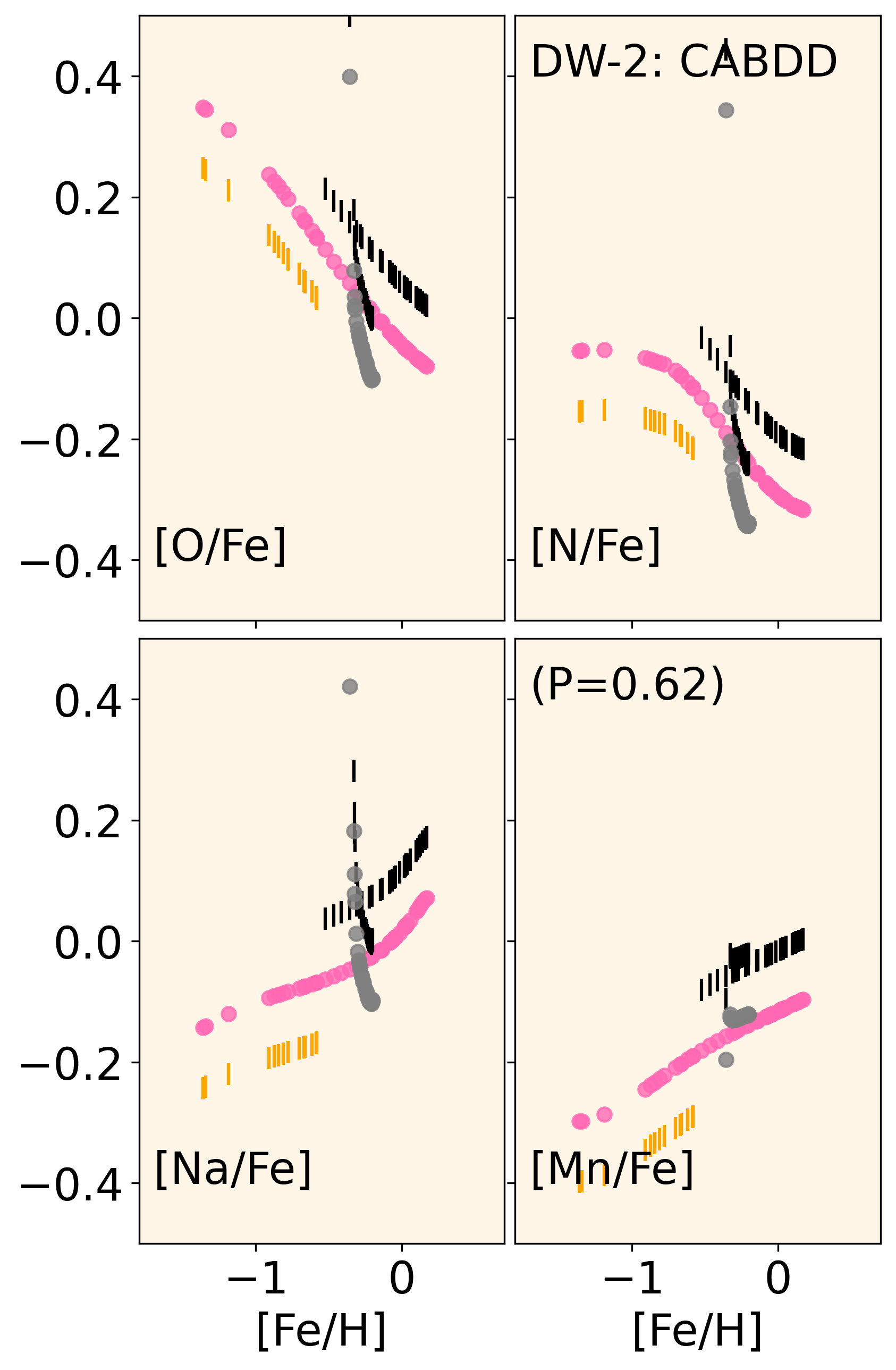}
\includegraphics[width=0.192\linewidth]{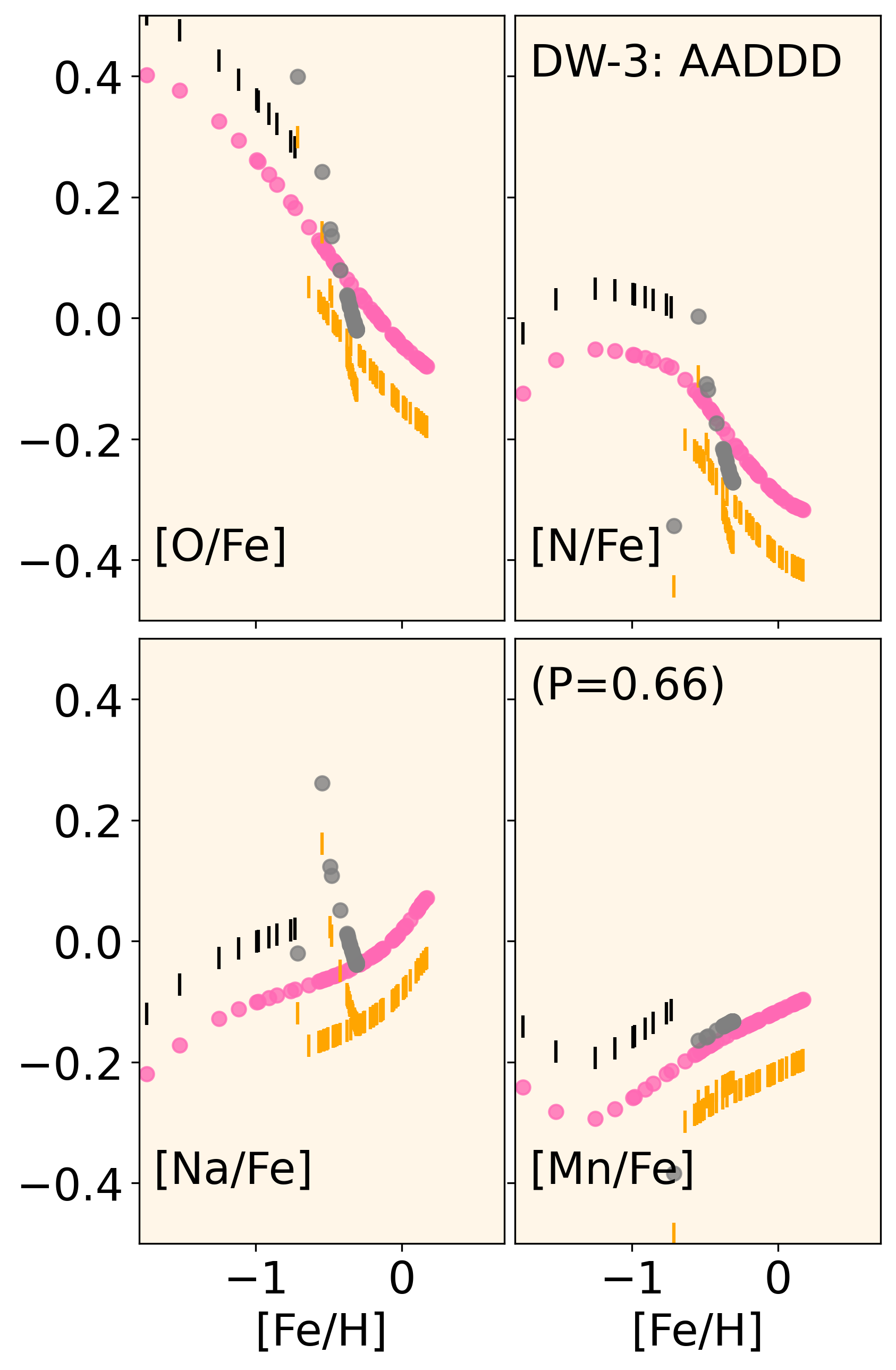}
\includegraphics[width=0.192\linewidth]{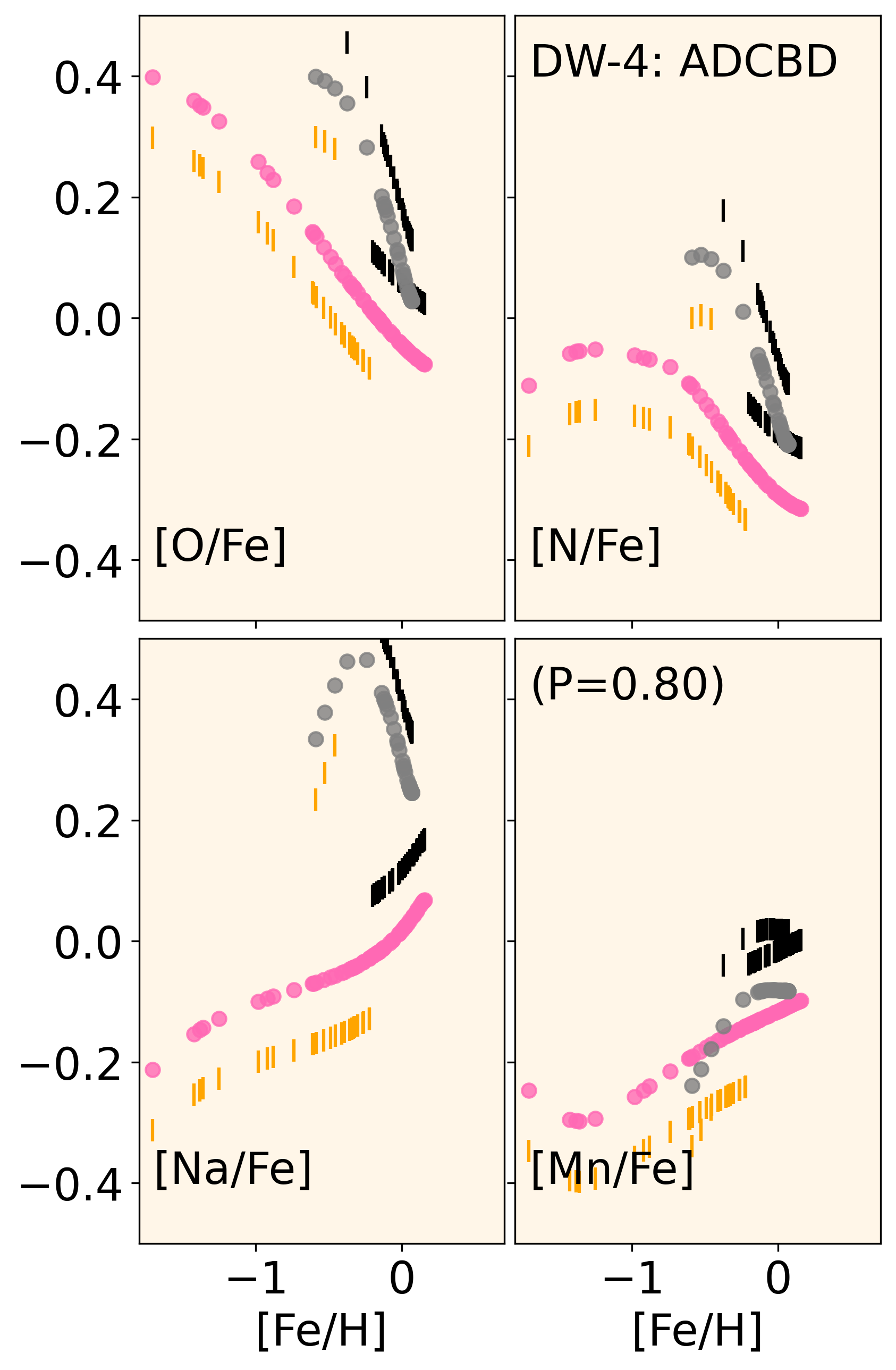}
\includegraphics[width=0.192\linewidth]{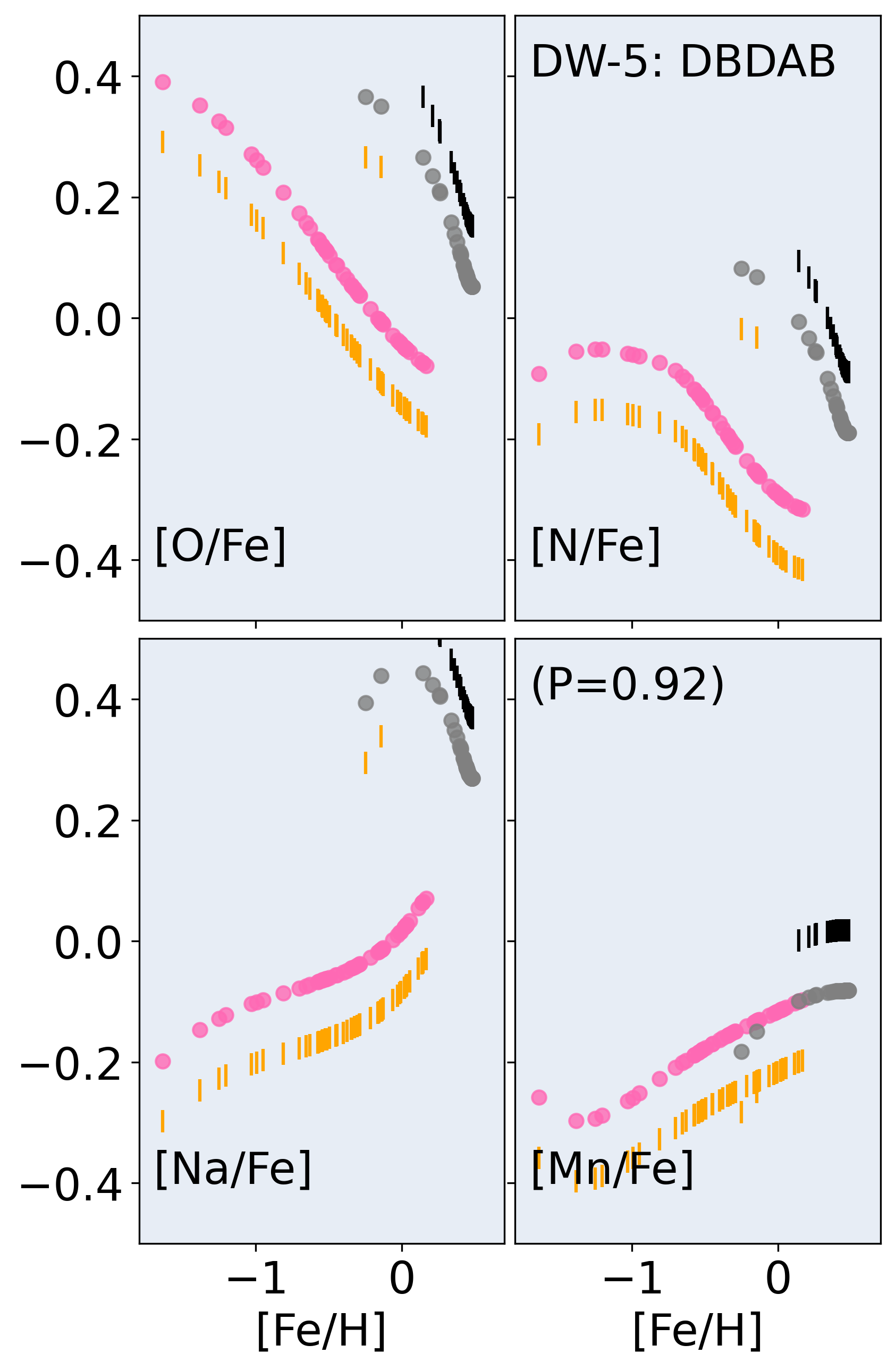}
\caption{Abundance planes for the example models shown in Fig.~\ref{fig:examples_015noise_100}. Green corresponds to the \mwfid\ model, pink to the \dwfid\ model and gray the grid model. Vertical lines correspond to the result of clustering in abundance scale using two clusters, one represented with black and the other with orange. }
\label{fig:planes_gmm}
\end{figure*}

We compare the purity of the NJ branches with the result of a clustering algorithm on the chemical abundances of the models discussed in Sect.~\ref{sect:cases}.  The clustering uses Gaussian mixture models (GMMs) and considers all abundances, in [X/H] form, used for the NJ tree building algorithm, to find two groups. We implemented the model from the Python package {\tt sklearn}. Fig.~\ref{fig:planes_gmm} shows a similar plot to Fig.~\ref{fig:planes} but includes the result of the GMM with overplotted vertical lines shifted in the Y axis by an arbitrary amount for better visualization. One cluster is shown in black and the other in orange. 

It is possible to see that cases with $P < 0.9$ (the NJ is unable to separate the models in distinct branches), the GMM does not perform a good classification of the groups. For the cases in blue, sometimes the GMM does a good classification (MW-1, MW-3, MW-5) but in other cases it clusters the data into wrong groups (DW-4 and DW-5). In fact, the GMM does a good job in classification when the abundance pathways of the models are tight. When the abundance pathways are extended in metallicity, the GMM separates metal-poor from metal-rich, regardless of the model. The NJ, on the contrary, can create branches with extended metallicity distributions.

To evaluate that these examples apply to the entire dataset, we perform a GMM clustering to the entire grid. To do so, we select 100 random data points from the combined models between a grid and a fiducial model, add an uncertainty of 0.15 dex to all abundances and perform the GMM clustering to find two groups. In this clustering, it is possible that one group is much larger than the other. Only for groups of comparable size, that is, when the ratio between the sizes of both groups is within 50\%, we calculated the purity following Eq.~\ref{eq:purity}. For other groups, we set the purity to zero. 

\begin{figure*}[t]
\centering
\includegraphics[scale=0.5]{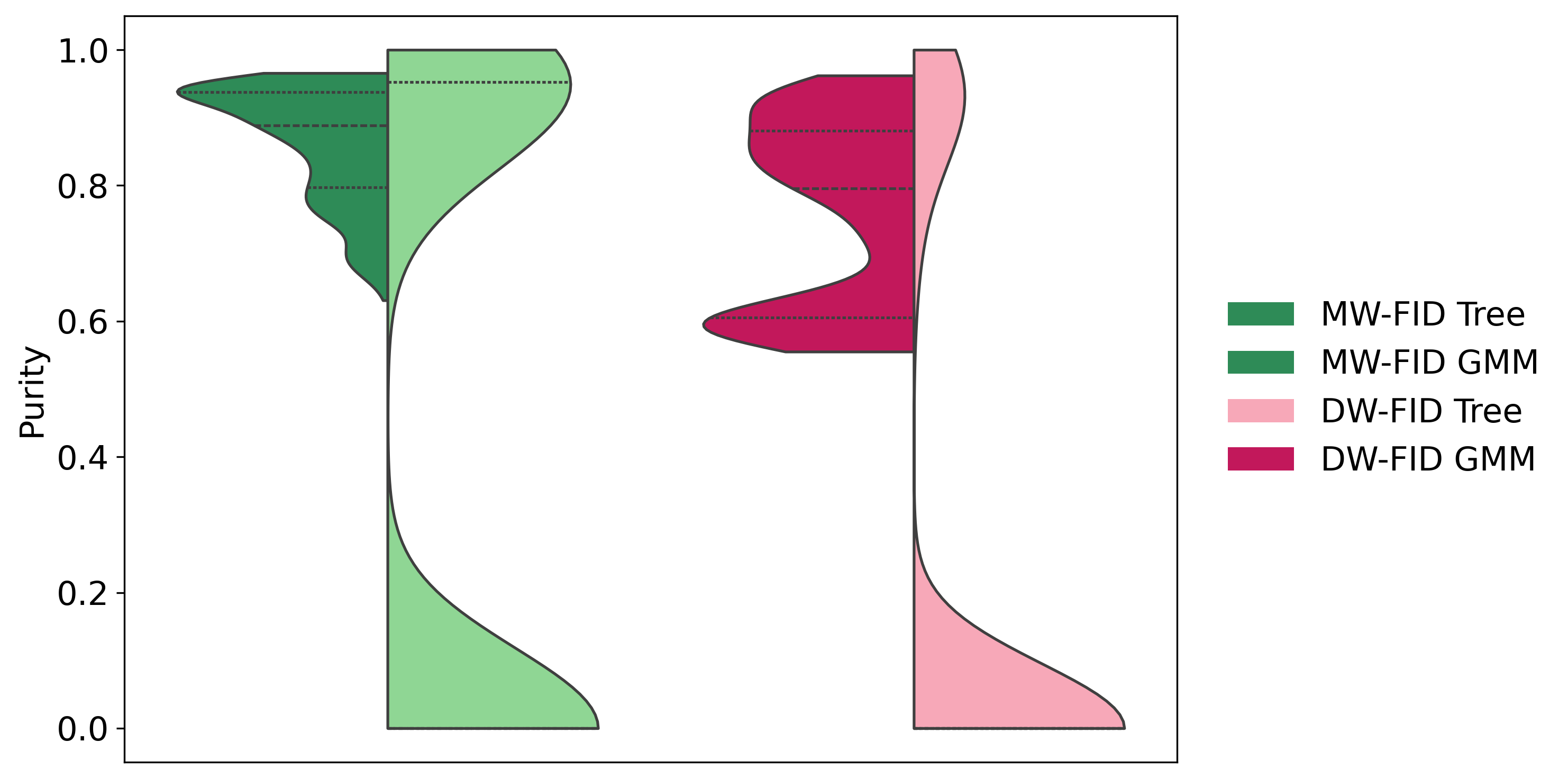}
       \caption{Violin plot for purity distributions for all models using Eq.~\ref{eq:purity} using the phylogenetic trees (Tree) and using the GMM classification.}
   \label{fig:violin}
\end{figure*}

Figure~\ref{fig:violin} shows the violin plots for all grid models combined with the \mwfid\ model in green and the \dwfid\ model in pink. Dark colors represent the distribution of purities calculated from the trees, while light colors represent the distribution of purities from the GMM.  It is possible to see that in a very large number of cases the GMM does not perform a good classification, especially for the \dwfid\ case.

\end{document}